\documentclass[final]{agujournal2019}
\usepackage[english]{babel}
\usepackage{newtxtext,newtxmath}
\usepackage{apacite}
\usepackage{natbib}
\usepackage[table]{xcolor}\definecolor{darkblue}{rgb}{0,0,0.5}\definecolor{firebrick}{rgb}{0.75,0.125,0.125}\definecolor{darkgreen}{rgb}{0,0.45,0}
\usepackage[colorlinks=true,linkcolor=firebrick,citecolor=darkgreen,urlcolor=darkblue]{hyperref}
\usepackage{booktabs}
\usepackage{siunitx}
\usepackage[capitalize]{cleveref}
\usepackage{wrapfig}
\usepackage{enumitem}
\draftfalse

\journalname{JGR: Atmospheres}

\begin{document}

\title{A Terrestrial Gamma-ray Flash with an Asymmetric Footprint Observed at the Pierre Auger Observatory}

\authors{
The Pierre Auger Collaboration,\affil{1}\thanks{Full list of authors of the Pierre Auger Collaboration can be found at the end of the document}~
John Ortberg,\affil{2}~
David M.~Smith,\affil{2},
and
Joseph Dwyer\affil{3}
}
\affiliation{1}{Observatorio Pierre Auger, Av.\ San Mart{\'\i}n Norte 304, 5613 Malarg\"ue, Argentina}
\affiliation{2}{Physics Department and Santa Cruz Institute for Particle Physics, University of California, Santa Cruz, California, USA}
\affiliation{3}{Department of Physics and Astronomy, University of New Hampshire, Durham, New Hampshire, USA}

\correspondingauthor{John Ortberg}{jortberg@ucsc.edu}
\correspondingauthor{Pierre Auger Collaboration}{spokespersons@auger.org}

\begin{keypoints}
\item We triangulated the gamma-ray source of a TGF seen by more than 40 detectors at the Pierre Auger Observatory to an altitude of \qty{2.3\pm0.5}{km} and horizontal uncertainty of \qty{\pm82}{m}.
\item The emission pattern of the TGF was not only asymmetric about its vertical axis, but it also had two distinct peaks in fluence offset by about \ang{120} in the azimuthal direction.
\item\textsc{Geant4} simulations suggest that the TGF likely had an emission distribution with a strong component above $\theta = \ang{75}$, the off-nadir angle.
These simulations also suggest a brightness of ${\geq}10^{16}$ photons above \qty{1}{MeV}.
\end{keypoints}

\begin{abstract}
The Pierre Auger Observatory, a \qty{3000}{km^2} detector array that sits \qty{1400}{m} above sea level in Argentina, has the ability to map the entire footprint of Terrestrial Gamma-ray Flashes (TGFs), which are short, intense bursts of gamma rays associated with lightning.
This is in contrast to the vast majority of TGF detections that only occur in a single detector in space.
In this paper we leverage this capability to perform two important analyses on an event from 2007, chosen for its large footprint and high signal quality.
First, we triangulate the source of the TGF to \qty{2.3\pm0.5}{km} height above ground level using the arrival time of the gamma rays themselves at the detector array, independently of any information regarding the lightning channel.
With the source position established, we can analyze the spatial distribution of the TGF on the ground with respect to that source.
Not only do we find that one side of the TGF footprint shows an order of magnitude higher flux than the other, but we also find an asymmetric azimuthal structure consisting of two clear peaks in intensity.
These peaks are offset by about \ang{120} from each other.
This azimuthal structure is not possible under the traditional uniform electric-field model of a TGF.
Using \textsc{Geant4} simulations of collimated beams of the relativistic runaway electron-avalanche spectrum, we attempt to reconstruct the source geometry of the TGF and find that it likely had significant components in the $\theta > \ang{75}$ range, where $\theta$ is the off-nadir angle.
\end{abstract}

\noindent
\textbf{Plain Language Summary}
\\
Terrestrial Gamma-ray Flashes (TGFs) are intense bursts of high-energy gamma rays induced by lightning flashes. They are bright enough to be seen from space despite lasting less than a millisecond.
Although there have been thousands seen by satellites, and tens to hundreds seen by detectors on the ground, it is extremely rare to have multiple detections of the same event from different locations.
The Pierre Auger Observatory, which spans \qty{3000}{km^2} and was designed to observe cosmic-ray showers, is serendipitously capable of detecting TGFs.
One such TGF was observed by 54 detectors simultaneously, which allows us to sample the entire ``footprint'' of the TGF instead of just one point in the beam.
Analysis of the footprint reveals that the TGF was not only asymmetric about its center axis, but also that it appears to have come from two separate source regions.

\section{Background \& Summary}

Terrestrial gamma-ray flashes (TGFs) are extremely short, intense bursts of gamma rays associated with lightning.
The gamma rays are driven by so-called relativistic runaway electron avalanches (RREA) that can occur in electric fields inside thunderstorms, and are distributed in energy according to the well-known RREA spectrum, $\mathrm{d}N/\mathrm{d}E\propto\exp(-E/E_0)/E$, where $E_0\approx\qty{7}{MeV}$ \citep{dwyer03limit,dwyersmith05}.
Estimates of their frequency range between about 1 in \num{10000} lightning flashes at the lower limit to just under 1 in 100 at the upper limit \citep{briggs2013,Albrechtsen2018,fabro19}.
It is worth noting that although TGFs have generally been thought to always occur with an associated lightning flash, recent observations suggest that some TGFs can occur in thunderclouds without a flash \citep{Zhang2021NbeTgf,OstgardALOFT2023}.
This relative rarity, sub-millisecond duration, and extremely high flux (that can exceed the dynamic range of many detectors) all contribute to making TGFs extremely challenging to observe.

Given the greater field of view, it is much more likely to see a TGF from a detector in orbit than one on the ground.
While tens of thousands of TGFs have been observed from orbit over the last few decades, there have never been more than three satellites actively recording TGFs at the same time.
Furthermore, the orbital positions of the satellites are not synchronized and any coincident detections of the same event are left largely to chance.
Almost all observations have been from a single point in space, save for a handful of events visible from two satellites instead of one \citep{Marisaldi20}.
The photon distribution in a TGF has many unknown variables -- total brightness, angular distribution, axis orientation, source altitude, etc.\ -- so that using a single observation to determine one of them requires assuming the value of the others.

Monte Carlo simulations in previous works have had some success in constraining beam widths (emission-cone angles) and source altitudes associated with satellite observations \citep{hazelton09,mailyan16,lindanger21}.
However, not only were there multiple combinations of compatible beam widths and source altitudes for many observations, but they were dependent on assumptions about intrinsic brightness and perfectly vertical orientation.
\citet{Mailyan2019} performed similar analysis but found many cases where a non-zero tilt gave the best fit, once other source parameters were contained to reasonable ranges. 

The ground detections of TGFs that have occurred have also been almost exclusively single point observations \citep[e.g.][]{dwyer04rocket,bowers17,wada18}.
\citet{Ortberg23} did an analysis similar to \citet{Mailyan2019} and \citet{lindanger21}, where simulations suggested the TGF must have had some combination of horizontal tilt, wider angular distribution, or abnormal intrinsic brightness to account for the flux in their respective detectors.
However, without more spatial resolution, they could not conclude which parameter(s) was(were) responsible.

More recently, there have been reports of TGFs being detected over ground-based cosmic-ray observatories.
Although the observable field of view of these arrays is much smaller than that of a satellite, the TGFs that do occur over them will be seen by dozens of detectors instead of just one.
The Telescope Array in Utah has seen several events that, although usually shorter and a few orders of magnitude dimmer than those seen from space, have had their entire footprint fall within the bounds of the array \citep{2017Abbasi,2020BelzUtahTAB}.
The Pierre Auger Observatory in Argentina has recorded about 20 suspected TGFs dating back to 2005 \citep{Schimassek_2022,Colalillo2022}.
Here we will focus on a particular event from that database and show for the first time clear evidence of asymmetric azimuthal structure in a TGF.

\section{Data Sources and Methods}

\subsection{The TGF event}

The Pierre Auger Observatory \citep{AugerBackground2015} in Argentina includes a \qty{3000}{km^2} array of particle detectors specifically designed to study ultra-high-energy cosmic rays via the detection of extensive air showers.
However, due to its climate, lack of light pollution, and its wide variety of sky-surveying instruments, it has turned out to be a lucrative place to study exotic phenomena in atmospheric electricity \citep{2017AugerTGFsColalillo,AugerElves2020,AugerAtmoElectricity2023}.
It features 1660 water-Cherenkov detectors (WCD, each containing \qty{12}{t} of purified water) spaced at intervals of \qty{1500}{m}, or less in some parts of the array, that are used to measure the energy flow at the ground carried by the flux of particles in the air showers generated by the primary cosmic rays.
Thanks to their depth (\qty{1.2}{m}, corresponding to 3.3 radiation lengths), the detectors are sensitive not only to muons, electrons, positrons, but also to gamma rays above \qty{1}{MeV} via pair production of electrons and positrons.
The signal from each WCD is read out by three photomultiplier tubes (PMT) digitized at \qty{40}{MHz} via a 10-bit Flash Analog-to-Digital Converter (FADC), providing \qty{25}{ns} time resolution.
The upper limit of the dynamic range in deposited energy is about \qty{5e11}{eV/\micro\second}, above which the signal is saturated.
Each PMT signal is tagged with GPS time stamps to an absolute time accuracy of $\approx\qty{12}{ns}$. 
The shower data are obtained through a hierarchical trigger system described in detail in \cite{ABRAHAM201029}.
In short, triggers from each WCD are sent to a central acquisition system that searches for time and spatial clustering.
When a coincidence is found between at least three stations, data from triggered detectors are collected. 
In addition to the trigger times, the data include for each PMT also a \qty{19.2}{\micro\second}-long (i.e., 768 time-bins) signal trace from the FADC.

\begin{figure}
   \centering
    \includegraphics[width=1\textwidth]{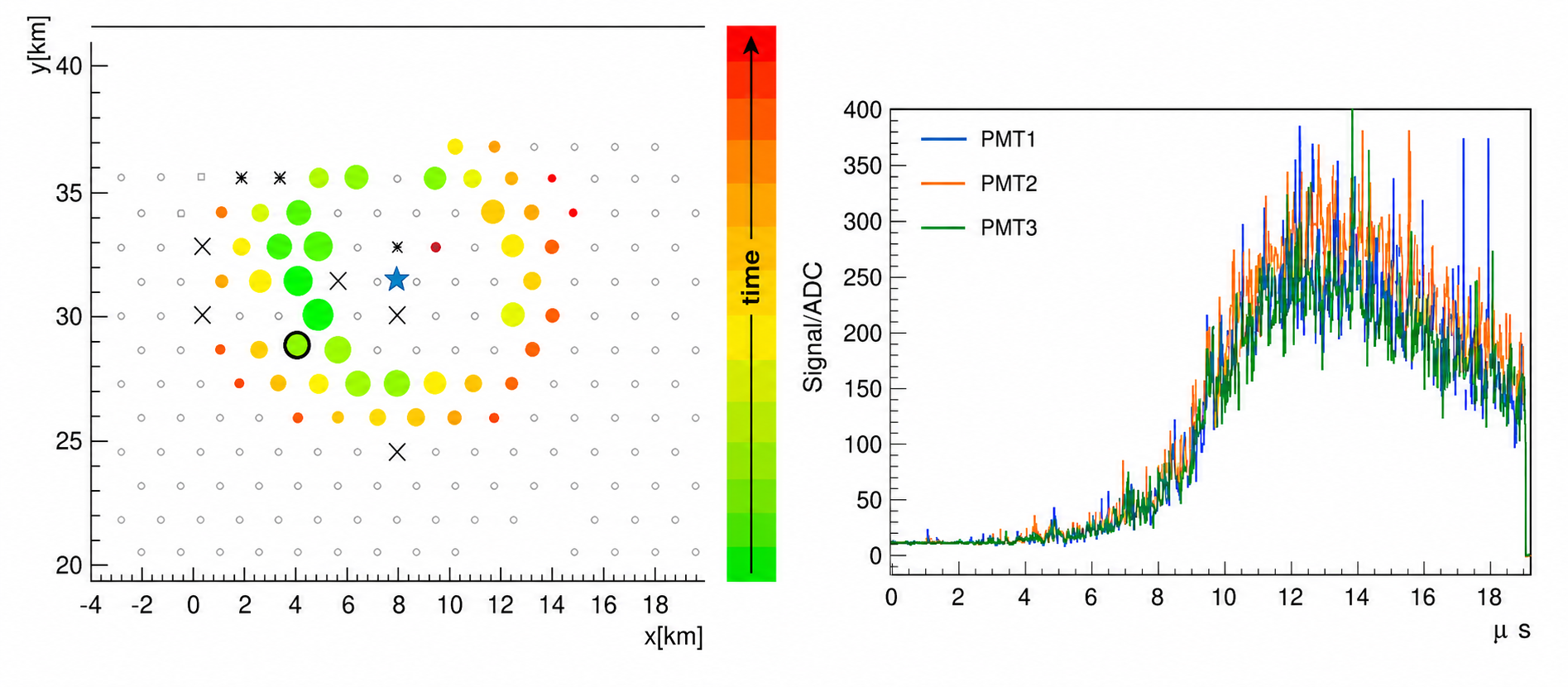} 
    \caption{The TGF event analyzed in this paper.
    \emph{Left}: Distribution of the triggered stations.
    The circular markers indicate the stations with the typical long signal of TGFs observed with, shown in the right panel.
    Their size is proportional to the amount of energy released in the station, while the color indicates the time at which gamma-rays reach it.
    The green stations are the first to receive the signal, while the red stations are the last.
    The stars represent the ``lightning stations'', while the crosses depict stations hit by atmospheric muons, which are detected by chance. The blue star represents the center of the footprint.
    \emph{Right}: Long and intense TGF signals.
    The three different colors represent the signals from the 3 photomultipliers (PMTs) of the station marked by a black circle in the left plot.}
\label{fig:TGFevent}
\end{figure}



Although the triggers have been designed to work optimally for air showers, they have also selected TGF events. 
\cref{fig:TGFevent} (left panel) shows the event that will be analyzed in this work.
Detected during a thunderstorm on 14 October 2007, 20:10 UTC, it has triggered 54 stations, covering an area of about \qty{150}{km^2}. 
As one can see, signals are missing in the central part of the footprint. 
In almost all TGF events, including this one, the nearest triggered stations are about 3 km from the center. 
However, the radial coverage is extensive, reaching 7 km here and up to 9 km in other cases; this allows for a large number of stations to be analyzed.
The missing signal in the center results from the combined effect of two factors. First, the station trigger is optimized for extensive air shower signals \citep{Colalillo2022}, which are shorter in duration and less intense. Second, during thunderstorms, the acquisition system might be overwhelmed by high-frequency electronic noise generating a large number of false triggers.

The signals in the triggered stations can be classified into three groups.
The most abundant includes the so-called ``long signals'', based on a comparison with shower signals.
The presence of stations featuring such signals is actually used to identify TGF events: they are represented in the left panel of \cref{fig:TGFevent} by colored circular markers, the size of which is proportional to the amount of energy released in the station, while the color is related to the arrival time of the gamma-rays.
The FADC signals read by the three PMTs of a station at \qty{5}{km} from the center (blue star in the right panel of \cref{fig:TGFevent}) of the footprint are illustrated in \cref{fig:TGFevent} (right panel). 
Each signal presents a slow rise, a peak and a long decay, so that the signal exceeds the \qty{19.2}{\micro\second}-long acquisition window.
The second group of signals is dubbed ``lightning'': three stations in the event, represented by asterisks in \cref{fig:TGFevent} left, show this type of signal, which is characterized by high-frequency electronic noise induced by atmospheric electricity caused by the thunderstorm.
This noise is also superimposed onto the long signals.
Finally, the third group of signals is due to atmospheric muons: stations with such signals are marked by crosses on the left \cref{fig:TGFevent} and are random coincidences with the event.
Other TGFs seen at the Observatory show similar characteristics; this event was chosen because it was the most energetic and therefore allowed the most accurate reconstruction.

\subsection{Geant4 Simulations}
\label{sec:GeantSimIntro}

\begin{figure}
    \centering
    \includegraphics[width=\linewidth]{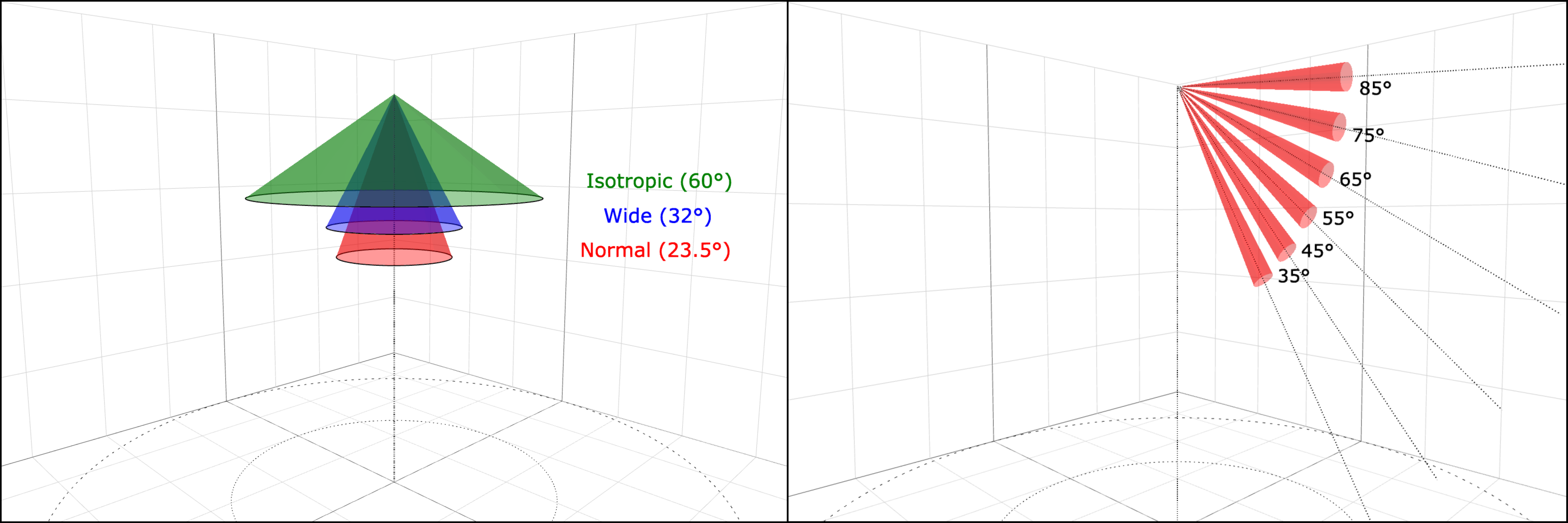}
    \caption{A schematic of the source geometries used in our simulations.
    \emph{Left:} The red, blue, and green cones visualize the half-opening angles for the normal, wide, and isotropic beams, respectively.
    These are the traditional azimuthally-symmetric and vertically-oriented TGF models.
    \emph{Right:} An illustration of how the monodirectional RREA beams were simulated at different nadir angles $\theta$, centered on $\phi = 0$.
    The results can then be convolved with an angular distribution function $h(\phi,\theta)$ to recreate a more complex source geometry.}
    \label{fig:SimulationSchematics}
\end{figure}

The goal of our simulations is to find the TGF source configuration that best predicts the fluence we observed at the ground.
This configuration will consist of a list of photons with specified position, energy, and direction, representing the final particles produced at the source.
Because TGF photon energies are largely constrained by the RREA spectrum regardless of mechanism or source geometry, our free parameter space comes from modifications to photon positions and angular distributions.
We use the photon list described in \citet{dwyer12tgftheory} as a starting point, and will refer to this as the ``normal'' beam configuration.
In that work, the simulated photons were produced inside the source region of a TGF with a vertically-directed, uniform electric field.
\citet{dwyer12tgftheory} points out that this uniform configuration leads to the narrowest possible angular distribution of photons, as any convergence or divergence in the field lines widens the overall distribution.
These angular distributions, or ``emission cones'', are often defined by the half-opening angle $\theta_\text{H}$ in which 50\% of the photons are contained \citep{gjest11}.
The source configurations and associated $\theta_\text{H}$ we will use, which are shown in \cref{fig:SimulationSchematics}, are as follows:
\begin{enumerate}
    \item Normal ($\theta_\text{H} = \ang{23.3}$): the photons produced by a TGF with a uniform RREA field as described in \citet{dwyer12tgftheory}.
    \item Wide ($\theta_\text{H} = \ang{32.9}$): the ``normal'' emission cone convolved with a Gaussian of width $\theta = \ang{25}$.
    \item Isotropic ($\theta_\text{H} = \ang{60}$): photon directions are completely isotropic in the downward hemisphere ($\theta\leq0$).
    \item Monodirectional ($\theta_\text{H} = \ang{0}$): photons are aligned in a perfectly collimated beam.
\end{enumerate}

Neither the isotropic nor monodirectional models are physical, but they still serve a purpose in our analysis.
The isotropic model represents the upper limit of ``widening'' the photons in an otherwise downward-directed electric field.
The monodirectional simulations allow us to consider non-standard and/or asymmetric distributions.
This can be performed, in principle, by approximating such a distribution as a combination or convolution of monodirectional beams at various azimuthal and nadir angles.
This approach would be severely underdetermined in events with only a single-point observation. 

The photons from the TGF were placed with the bottom of the avalanche region at \qty{2}{km} above ground level.
Although the avalanche region is about \qty{1}{km} in height, most of the photons are created within a few \qty{100}{m} of the bottom for an effective height of about \qty{2.1}{km}.
\citet{Ortberg23} showed that for large horizontal distances ($\geq\qty{3}{km})$, raising the altitude of a TGF decreases fluence on the ground. 
This is because, at those distances, any increase in intensity as a result of being more directly in the beam is offset by the gammas having to travel through more atmosphere. 

Once the source geometries of the gamma rays are established, we used CERN's \textsc{Geant4} (GEometry ANd Tracking) software for particle tracking \citep{AgostinelliGeant4,geantCite2,geant2016}.
Inside the \textsc{Geant4} model, we constructed an atmosphere according to \citet{UsStdAtm76}, which gives a globally-averaged vertical profile of temperature, pressure, and air density among other parameters.
We use the \textsc{Geant4} standard electromagnetic-physics packages known as \texttt{EM Option 0}, which prioritizes performance on large data sets over accuracy at small scales or high energies.
While this aligns with our goal of trying to map the entire footprint of a TGF, it is important to note that this approach ignores photonuclear reactions, which would be important to simulate if interest is in the TGF afterglow such as in \citet{enoto17} or \citet{bowers17}.

Due to the need to simulate flux at large angles from the TGF axis, we used a Monte-Carlo method with importance resampling to focus on the 3 to \qty{7}{km} range.
In the first stage, photons were simulated from the source out to the surface of a cylinder \qty{2}{km} away.
Any particles with energy less than \qty{1}{MeV} reaching the surface were removed in the first stage, and the remaining particles were in the second stage multiplied by a factor of 100.
Thus, an intensity of \num{e10} photons in the first phase actually represents a source intensity of \num{e12} photons in the second stage while suitably saving the computing time.

\section{Characterization of the TGF Event and its Source}

\subsection{Gamma-Ray Source Triangulation}
\label{sec:triangulation}

With the high spatial and temporal resolution of the surface detectors at the Observatory, we have the rare opportunity to triangulate the source of the gamma rays themselves; most TGF localizations are actually done to the tip of the lightning leader seen in the radio spectrum at the inferred time at the source. 
For the purposes of triangulation we treated the TGF as a point source. Though the source region may be hundreds of meters in diameter, the spatiotemporal evolution of the TGF source itself is likely to resemble a point source as the electric field enhancement from the lightning leader tip propagates outward at $c$.

\begin{figure}
    \centering
    \includegraphics[width=0.8\linewidth]{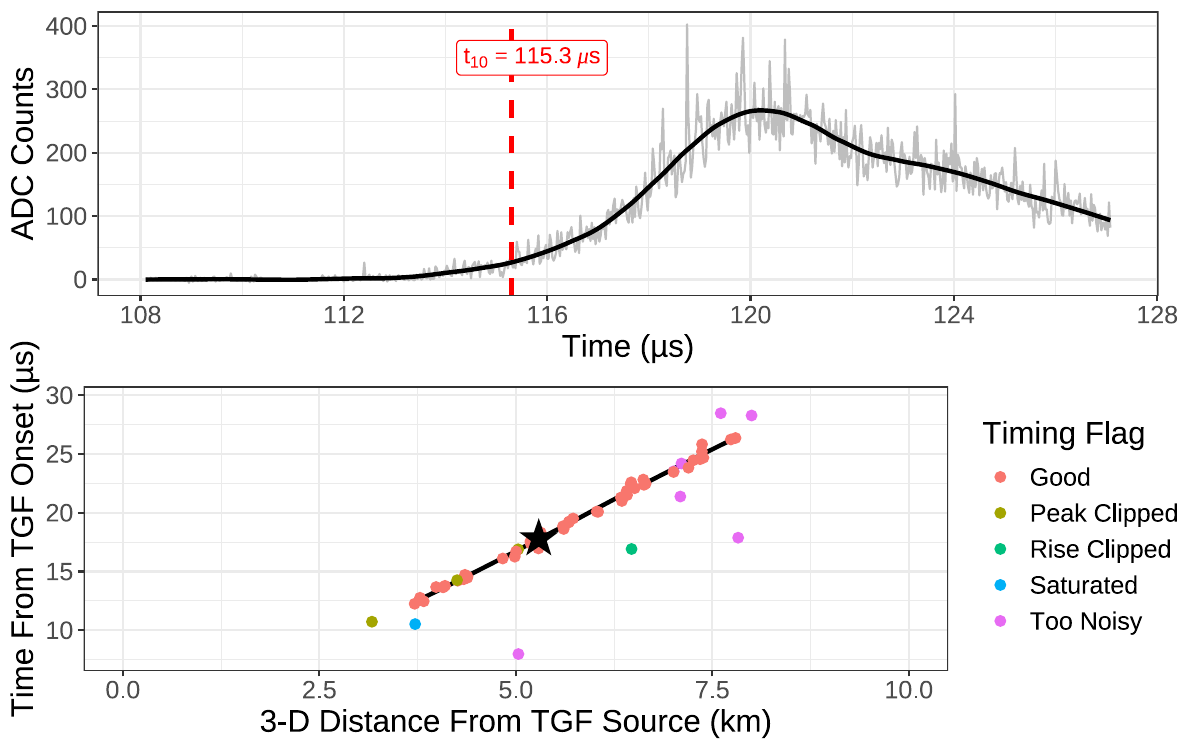}
    \caption{\emph{Top:} PMT signal from a surface detector that was located about \qty{5}{km} away from the TGF center.
    LOESS smoothing (black) is used on the raw signal (gray).
    We use 10\% of the smoothed-signal peak, which is reached at the red line, to compare arrival times between stations.
    The calculated TGF onset time is used to establish $t=0$.
    \emph{Bottom:} Photon time of arrival vs.\ distance from the TGF source.
    Rejected stations are also shown, as well as their reason for rejection.
    The time of arrival of the station in the top plot is shown with the black star $\bigstar$.
    A regression line (coefficient of determination $R^2=0.994$) is shown for the remaining stations, which also adds a first-order correction for atmospheric propagation effects at largest distances.}
    \label{fig:pmt}
\end{figure}

The detectors generally recorded a gradual rise in signal with various amounts of noise, so we implemented two strategies to determine a robust time of arrival at each station.
First, we used a locally weighted regression model \citep{LOESS1979} to smooth the PMT signal, and identified the time $t_{10}$ at which the smoothed signal reached 10\% of its maximum value.
Although the definition of an exact ``onset time'' is somewhat arbitrary, the 10\% fraction was chosen as a compromise for two reasons: (1) the fraction was high enough above the noise threshold of most stations to be coherent in the smooth signal and (2) $t_{10}$ was early enough in the time profile to be minimally affected by the pulse broadening caused by atmospheric scattering.
Choosing any other value from $t_5$ to $t_{50}$ only changed the triangulation result on average by \qty{5}{m} horizontally and \qty{200}{m} vertically.
This process is shown in the top panel of \cref{fig:pmt}. 

To account for the delay of $t_{10}$ that results from the pulse broadening mentioned above, we applied a first-order correction $\mu$ to the signal arrival time $t_\text{arr}$ based on \textsc{Geant4} simulations, i.e.,\ $t_\text{arr} = d/c - \mu d^2$ where $d$ is the distance from the source and $c$ is the speed of light.
We found $\mu$ to range from 0.0035 to \qty{0.0065}{\micro\second\per\square\kilo\meter} depending on the angular distribution used in the model, and used the average of \qty{0.005}{\micro\second\per\square\kilo\meter} for determining the TGF center.
Simulations also showed that asymmetry in the angular distribution had a negligible effect on the value of $t_{10}$ compared to a symmetric distribution. 

Second, we determined and eliminated any cases where the 10\% signal fraction could not be reliably calculated for one of four reasons (where the color of the corresponding data points in bottom \cref{fig:pmt} is added in parentheses):
\begin{enumerate}
    \item The apparent peak of the signal was closer than \qty{1.5}{\micro\second} of the end of the acquisition window (yellow).
    \item The signal was already rising at the onset of the acquisition window, therefore a baseline could not be established (green).
    \item The signal reached saturation of the FADC range, and therefore the true peak value is not known (blue).
    \item High signal-to-noise ratio: when the RMS of the residual between the raw signal and the smoothed curve exceeds 50\% of the peak value of smoothed curve (purple).
\end{enumerate}
Once a time of arrival $t_{10}$ is established for each station, we use a nonlinear regression on the equation mentioned above, $t_{arr} = t_{10} - t_0 = d/c - \mu d^2$, to triangulate the TGF source. 
The regression then solves for the spacetime coordinates of the TGF point source ($x_0, y_0, z_0, t_0)$. 
The three spatial coordinates $(x_0, y_0, z_0)$ are implicit in the calculation for the distance $d$ between each station and the TGF source.
The results of this regression are shown in the bottom panel of \cref{fig:pmt}, where the distance vs.\ time for each station is calculated from the final triangulated TGF source. 

Since the surface detectors are distributed on an almost perfectly flat, horizontal plane \citep{auger_dipole_paper}, the horizontal uncertainty is much lower than the vertical uncertainty; the 95\% CI is \qty{\pm82}{m} in the xy plane, \qty{\pm470}{m} in altitude, and $\pm\qty{0.49}{\micro\second}$ in onset time.
These uncertainties are calculated on the variability of the data alone, as, by comparison, the \qty{12.5}{ns} time resolution of the PMT has a negligible effect.
The source altitude, which was found to be \qty{2.3\pm0.47}{km} above ground level (i.e.\ \qty{3.7}{km} above sea level) is not uncommon among other downward TGFs detected from the ground \citep{abbasi18, wada19, Smith18, Ortberg23}.


\subsection{Extent of the Beam and Asymmetry}
\label{sec:AziResults}

\begin{figure}
    \centering
    \includegraphics[width=0.8\linewidth]{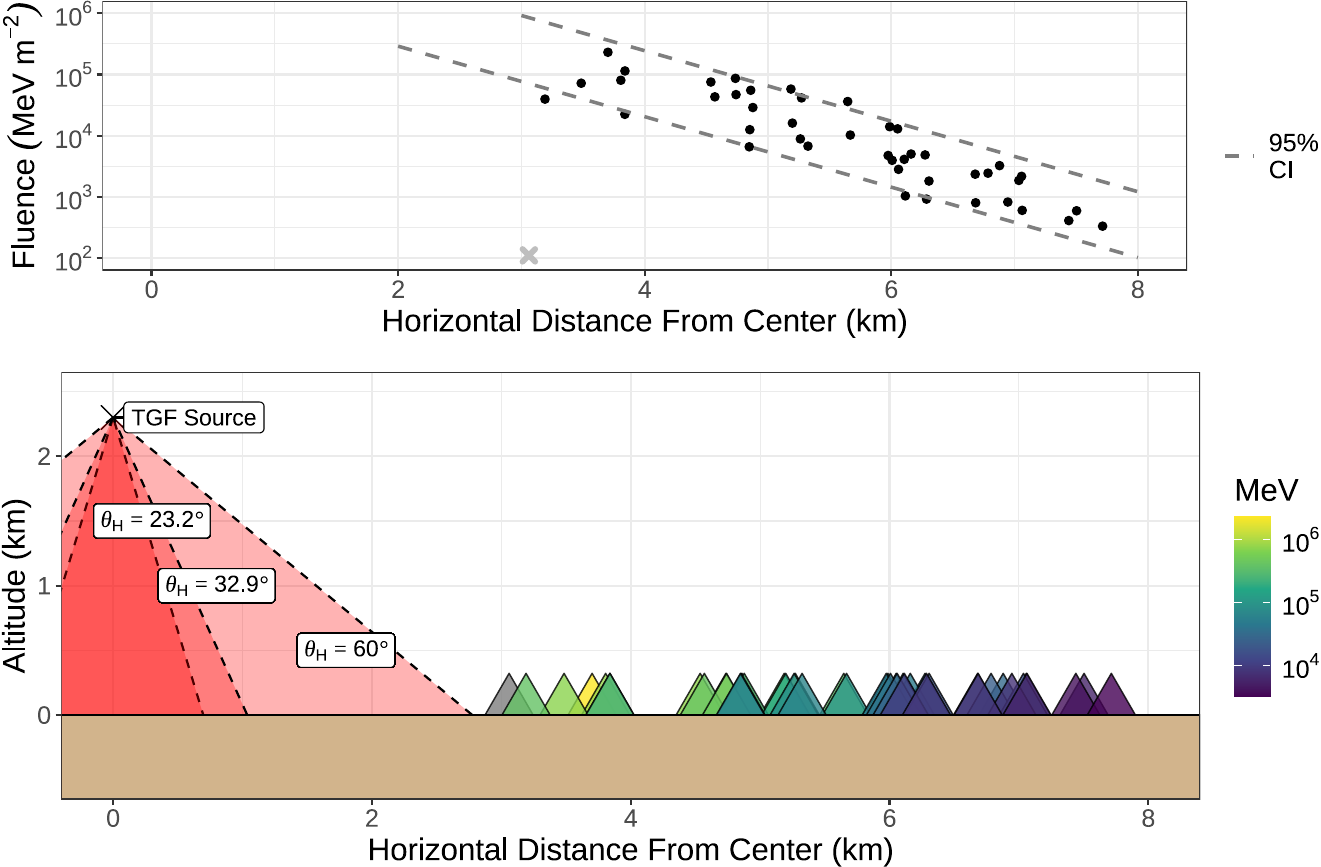}
    \caption{\emph{Top:} Fluence received by each detector as a function of \emph{horizontal} distance from the TGF source. 95\% confidence intervals are shown, and a station rejected due to saturation is shown as a grey $\times$.
    \emph{Bottom:} A to-scale illustration of the stations receiving signal relative to the low altitude TGF source.
    The red cone illustrates the half-opening angles $\theta_\text{H}$ as in \cref{fig:SimulationSchematics} -- it is unusual to see so many observations far away from the half-opening cone.}
    \label{fig:BrightnessModel}
\end{figure}

The calculation of the total energy deposited in each detector was carried out in two steps.
First, we converted the signal traces recorded by the three PMTs from the unit of ADC-counts to the reference unit of a vertical equivalent muon (VEM), determined for each detector using atmospheric muons \citep{2006SDCalibration}.
\qty{1}{VEM} is the charge associated with a vertical muon passing through the center of the detector and corresponds to an energy deposit of ${\sim}\qty{240}{MeV}$.
We then applied an asymmetric Gaussian fit to the converted traces.
After performing an unweighted chi-square minimization, three quality cuts were applied:
\begin{enumerate} 
\item The Gaussian peak must be located between 2.5 and \qty{17}{\micro\second}, ensuring that a substantial portion of both the rising and falling edges of the signal were detected. 
\item To ensure that the fit accurately reproduces the portion of the trace within the acquisition window, we required that
\begin{align}
\left|{\textstyle\int_0^{768}f(x)\,\mathrm{d}x} - I_\text{tot}\right| < 0.05 \, I_\text{tot},
\end{align}
where $f$ represents the best-fit function and $I_\text{tot}=\sum_{i=0}^{768}\text{trace}[i]$.
\item Finally, we required the difference between the integral of the best-fit function -- calculated over a time interval significantly wider than the acquisition window -- and the total trace content be below a threshold determined by a detailed study of successful and failed fits. 
This cut is crucial because, in certain cases, the constraint described in point 1 is insufficient to properly bound the falling edge of the signal, which could lead to a significant overestimation of the signal.
\end{enumerate}

For a station to be included in the following analysis, the fit must pass the quality cuts for at least two PMTs. 
The fluence is calculated as the average of the best-fit function integral, which is then divided by the station area to obtain the fluence per square meter.
Note that these selection criteria are more stringent than those used for the fit procedure for the triangulation: while the latter requires only precise definition of the peak position to determine the start-time of the trace, the reconstruction of the energy deposit also requires that a significant portion of the signal be contained within the acquisition window.
Of the 54 triggered stations, 43 could be included to determine the energy deposit versus 47 in the source triangulation (39 met both sets of criteria).
The fluence per square meter for the 43 stations that passed the cuts is shown in the top panel of \cref{fig:BrightnessModel} as a function of their distance from the center. Note that although arrival times were plotted as a function of \textit{total} distance from the source in \cref{fig:pmt}, fluence is now being plotted vs.\ horizontal distance from the vertical TGF axis.

The bottom panel of \cref{fig:BrightnessModel} shows a to-scale side profile of the radial and vertical distances of the stations to the triangulated source.
The half-opening angle of a normal width TGF beam, (see \cref{sec:GeantSimIntro}) shown in red, intersects the ground less than \qty{1}{km} from the source axis.
This is in sharp contrast to the farthest surface detector with a detectable signal at \qty{7.7}{km} away, which is a \qty{73.4\pm2}{\degree} offset accounting for the triangulation uncertainty.
This is among the most extreme observation angles reported, and poses challenges to the traditional model of a TGF as RREAs in a roughly uniform, vertical electric field.

Most TGF detecting satellites report being sensitive to TGFs up to a maximum of \ang{60} off-nadir \citep{briggs2013,Albrechtsen2018,ostgaard19}.
Downward TGFs seen from the ground are generally seen below that angle \citep{hare16,Smith18,Wada2020,Chaffin2024}, although one observation was made at an estimated \ang{73} \citep{Ortberg23}.
\citet{Ortberg23} suggests one possible explanation of their observation from a \ang{73} offset, i.e., that the entire TGF may have been oriented toward their detector, but in this case we see far reaching flux at a large range of azimuthal angles. 


\begin{figure}
    \centering
    \includegraphics[width=0.8\linewidth]{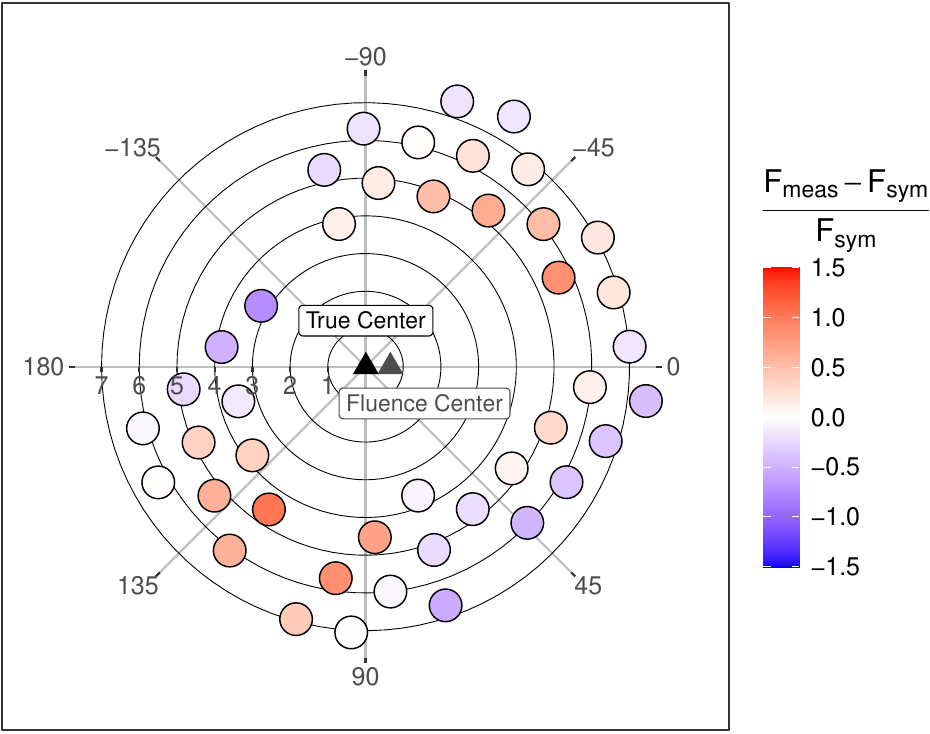}
    \caption{Measured fluence, $F_\text{meas}$ is compared to a symmetric best fit where $\log(\text{fluence}) \propto \text{distance}$, $F_\text{sym}$; red indicates stations that exceeded the fluence expected from a distribution symmetric in $\phi$, blue indicates those that showed a relative deficit.
    The center of the TGF, if modeled as the weighted average position of the fluence, is offset by about \qty{600}{m} from the true center calculated by photon arrival time, as shown by the triangles on the plot.
    The coordinate system has been rotated about the true center so that the fluence center lies on the x-axis at $\phi=\ang{0}$}
    \label{fig:AsymOverview}
\end{figure}

\begin{figure}
    \centering
    \includegraphics[width=\linewidth]{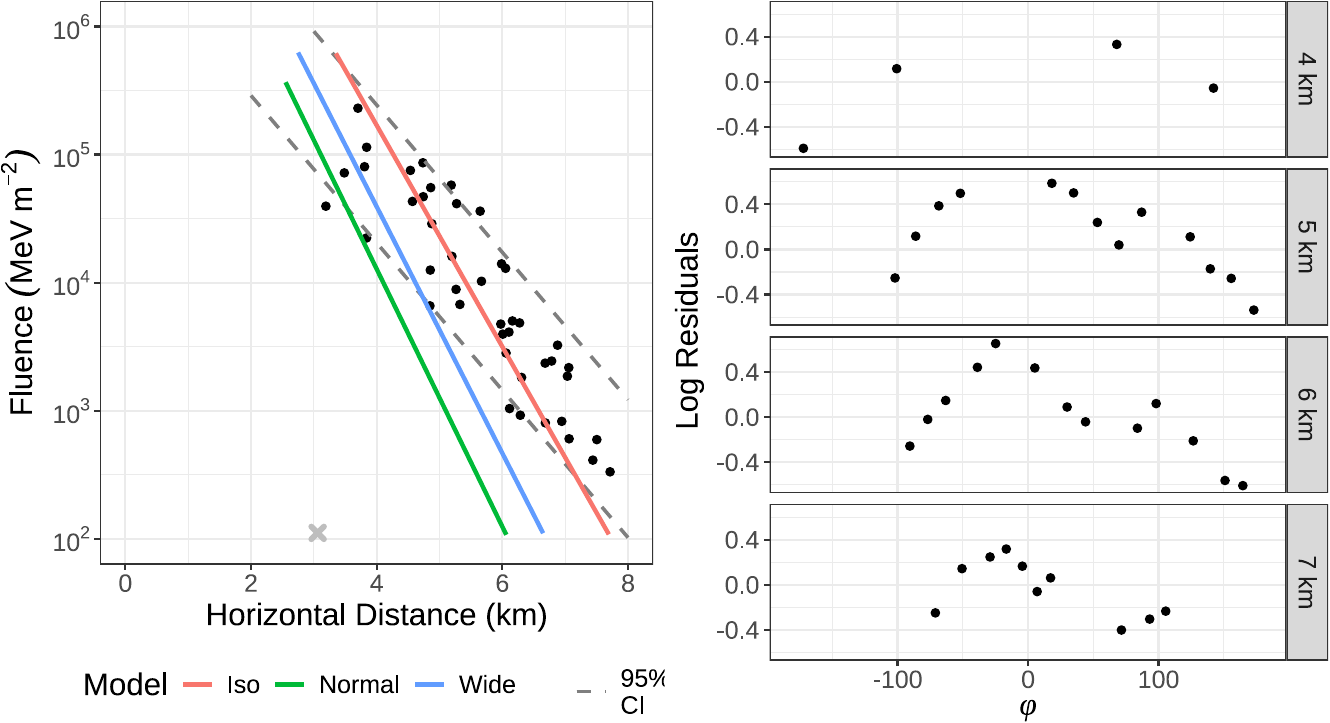}
    \caption{\emph{Left:} The same observations of fluence vs.\ distance as in \cref{fig:BrightnessModel} are plotted against simulation results of three different beam widths at standard (\num{e17} gammas) brightness.
    All three of these symmetric beam types appear to overestimate the slope of the fluence decay over distance.
    \emph{Right:} Residuals from the best fits in the left plot are shown as a function of azimuthal angle.
    In addition to a peak just left of $\phi=\ang{0}$, a possible secondary peak appears near $\phi=\ang{100}$.}
    \label{fig:Residuals}
    \label{fig:tradModels}
\end{figure}

The top down view of all stations in \cref{fig:AsymOverview} shows a clear asymmetry on the ground about the vertical axis of the TGF.
Such an asymmetry is the speculated outcome of a TGF tilted on its axis of emission while still being azimuthally symmetric about that axis.
\citet{Berge2019DownwardSims} found a method to quantify the relationship between the fluence distribution on the ground and the tilt and width of the TGF source based on simulations.
Using their methodology on our data, we find that our results do not fit any of their simulation outputs; the beam appears to be more tilted and more diffuse than even their greatest tilt angle (\ang{30}) and widest beam distribution ($\theta_\text{H} = \ang{45}$).
This is one motivation for an investigation of alternative models. 


A second motivation to investigate alternative models is the large unexplained variance on the leftmost plot of \cref{fig:Residuals}. 
Here, fluence as a function of distance nicely follows an exponential-decay trendline (i.e., $\log(\text{fluence}) \propto \text{distance}$), but the 95\% confidence intervals span more than an order of magnitude despite measurement uncertainty being negligible on this scale (${<}2\%$). 
Thus, on the right we plot the residuals with respect to the best fit line as a function of azimuthal angle, with stations grouped into bins by their distance from the center, rounded to the nearest \unit{km}. 
In addition to a strong peak near $\phi = -\ang{20}$, there is a clear lack of bilateral symmetry about that angle, with the emergence of a possible second smaller peak near $\phi = \ang{100}$. 
While a tilted TGF that is symmetric about its emission axis could be asymmetric about the azimuthal angle on the ground, it would still \emph{necessarily} have bilateral symmetry about some angle $\phi$. 
This suggests that we should explore a model that is neither symmetric about the vertical axis nor its own emission axis.

\section{Interpretation of the Results}
\label{sec:Interpretation}

\subsection{Comparison with Traditional Models}

The output of the \textsc{Geant4} simulations for each of the three traditional TGF models, as specified in \cref{sec:GeantSimIntro}, are plotted against observed data in the left panel of \cref{fig:tradModels}. 
Although the simulations were run with \num{e12} photons at the source, all three models are shown at a brightness of \num{e17} by simply scaling the output by a factor of \num{e5}.
This was done because \num{e17} gammas above \qty{1}{MeV} at the source is considered to be the ``standard intrinsic brightness'' by the majority of theory, observations, and modeling of TGFs seen from space \citep{dwyersmith05,celestin11,dwyer12tgftheory,Mailyan2019,lindanger21}. 

Different scaling factors can be chosen to find a better fit to the data outside of the standard \num{e17} brightness. These scaling factors will simply translate the output curve up or down proportionally on a semilog plot, without affecting its slope.
The best-fit brightnesses for normal, wide, and isotropic models are $10^{18.7}$, $10^{18.2}$, and $10^{17.4}$ gammas, respectively (scaling factors of $10^{6.7}$, $10^{6.2}$, and $10^{5.4}$).
Brighter than standard TGFs, on the order of \num{e18} to \num{e19} photons, are thought to be possible, but accepting such a model should be done with some skepticism.
Moreover, adjusting the brightness does not change the shape or slope of the curve, and all three models have a spatial decay that is significantly sharper than data.
A normal model best-fit brightness, for example, still underestimates the fluence in distant (${>}\qty{7}{km}$) stations by an order of magnitude while overestimating that in close stations by the same amount.
Even if we were to use the best fit line from \cref{fig:Residuals}, which is not congruent with any of the traditional models, the residuals still span more than an order of magnitude.
All of these factors suggest we should attempt to investigate the observed fluence distribution with alternative models.


\subsection{Constructing an Alternative Model}

\begin{figure}
    \centering
    \includegraphics[width=\linewidth]{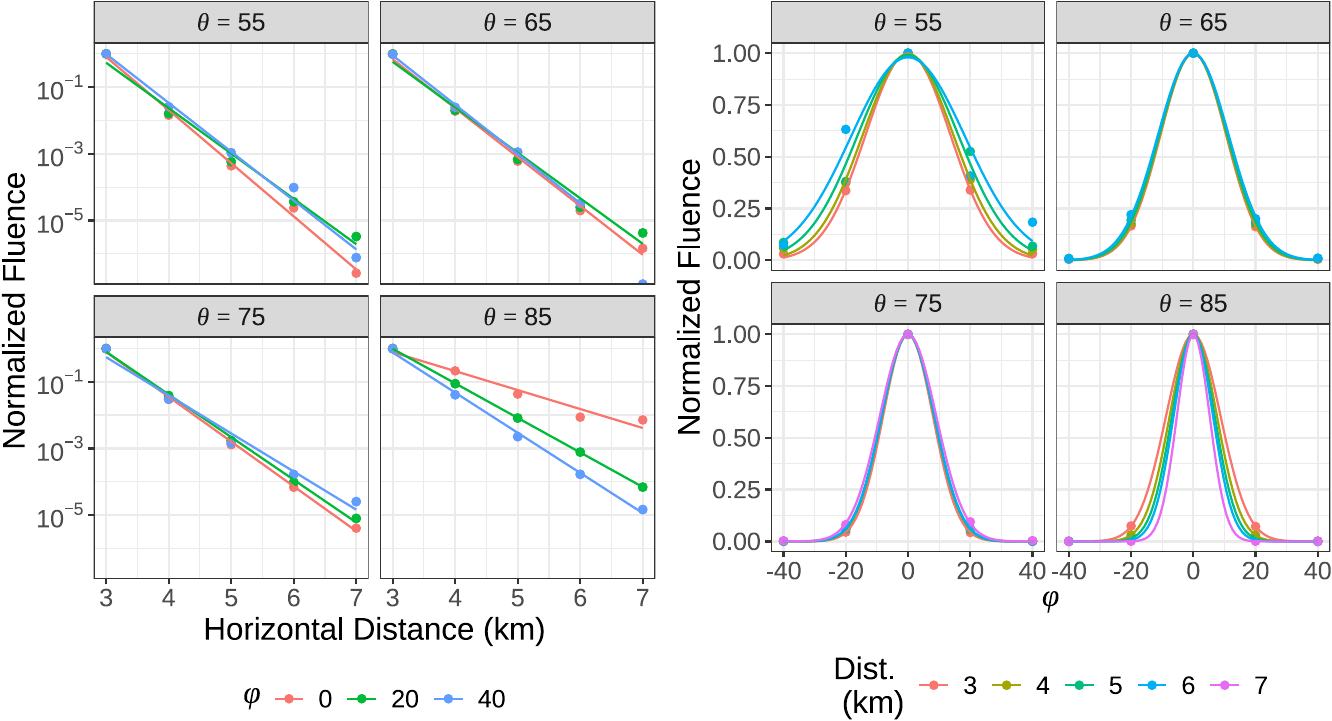}
    \caption{Results of \textsc{Geant4} simulations, collimated beams at RREA spectrum.
    Lines represent an exponential decay fit as a function of distance (left panel), and a Gaussian fit as a function of azimuthal angle (right panel).
    $\theta$ is off-nadir angle and $\phi$ is the azimuth where \ang{0} is along the axis of the beam.
    Distance is measured horizontally from the end of the avalanche region, considered the ``center'' of the TGF.
    Due to its collimated emission, any spread in $\phi$ is solely due to atmospheric scattering.}
    \label{fig:simulations}
\end{figure}

In order to construct an alternative model, we will attempt to recreate the observed fluence with a linear combination of monodirectional RREA beams.
In these monodirectional simulations, individual RREA beams were aimed at various off-nadir angles, $\theta_i$, and fluence was captured as a function of $r$ and $\phi$.
To prevent ambiguity, we will denote the ground-position azimuth by $\phi$ and the source-emission azimuth by $\phi_\text{s}$.

While the atmospheric response for a monodirectional beam differs for each angle $\theta$, it is rotationally invariant with $\phi_\text{s}$.
Therefore it is sufficient to simulate each beam on $\phi_\text{s} = \ang{0}$ and assume the same atmospheric response for all $\phi_\text{s}$.
Each monodirectional beam is effectively a delta function, i.e.\ $s_{\theta_i}(\phi_\text{s}) = \delta(\phi_\text{s})$, where $s_{\theta_i}(\phi_\text{s})$ represents the source distribution for a particular $\theta_i$.
It's then possible to convolve any source distribution, $s_{\theta_i}(\phi_\text{s})$, with the simulated atmospheric response, which we'll call $f_{\theta_i}(r, \phi)$, to get the fluence on the ground. That convolution, in principle, is
\begin{align}
    F(r,\phi) = (f_{\theta_i} * s_{\theta_i})(\phi).
\label{eq:convolution}
\end{align} 
Any distribution in $\theta$ at the source, however, must be reconstructed using a linear combination of the discrete values of $\theta_i$ done in the simulation.
Although we will take a simpler approach, theoretically the fluence of any source distribution $s(\theta, \phi)$ can then be approximately reconstructed with $F(r, \phi) = \sum_i K_i \int^{2\pi}_0 f_{\theta_i}(r, \phi - \phi_\text{s})\,s(\theta_i, \phi_\text{s})\,\mathrm{d}\phi_\text{s}$.

Results from these monodirectional simulations are shown in \cref{fig:simulations}.
For each data series, the fluence is normalized so that the maximum value is equal to 1 to allow for comparison of distributions.
Several combinations of $r$, $\theta$, and $\phi$ did not receive enough counts to produce meaningful statistics and were thus omitted from the plots.
Since our simulation was sensitive to fluence values of \qty{e-1}{MeV/m^2} assuming standard brightness, and the lowest recorded signal was \qty{3.32e2}{MeV/m^2}, it is likely that those lack of counts still tell us meaningful information about our TGF source.
The lack of counts for $\theta < \ang{55}$ suggests that the source likely had strong components at $\theta \geq \ang{55}$.
The lack of counts at $\phi > \ang{40}$ suggests that the signal at any one station is mostly due to photons originally aimed within \ang{40} of that station. 

The plot on the right of \cref{fig:simulations} shows that, although the fluence decreases with radial distance (as seen in the plot on the left), the shape of its distribution in $\phi$ varies little in the area of interest ($\geq\qty{3}{km})$.
This, along with the fact that introducing an $r$-$\phi$ coupling would likely overcomplicate and underconstrain the model, motivates our attempt to fit the data to a function that is separable in $r$ and $\phi$, i.e.
\begin{align}
    F_\text{obs}(r, \phi) \approx R(r) \, P(\phi)
\end{align}
where $F$ is the fluence distribution on the ground, and $R$ and $P$ are as of yet unknown functions of a single variable that multiply to produce the fluence.
Modeling the \emph{data} as smooth functions in $r$ and $\phi$ will also allow us to estimate a source distribution by deconvolving those functions with the atmospheric response generated by the simulations. 

\begin{figure}
    \centering
    \includegraphics[width=\linewidth]{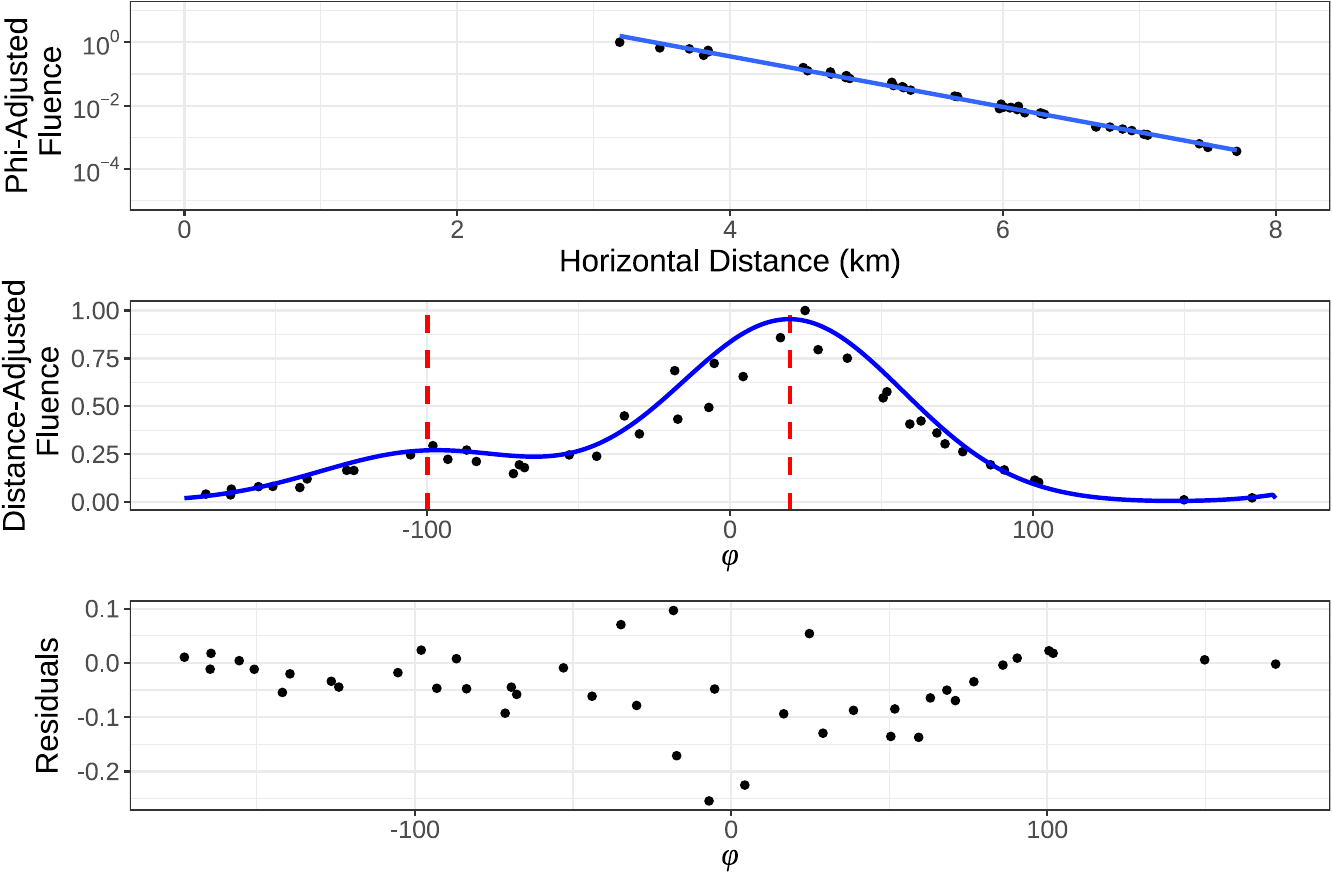}
    \caption{Final results of recorded fluence vs.\ distance (\emph{top}) and azimuthal angle (\emph{middle}).
    Fluence numbers are relative, accounting for the modeled effect of the variable \emph{not} on that plot.
    In other words, distance-adjusted fluence is $F_{obs} / R(r)$ and phi-adjusted fluence is $F_{obs} / P(\phi)$.
    The blue lines represent the two separable parts of the model described in \cref{eq:fullmodel}.
    Residuals (absolute) are shown as a function of $\phi$ in the bottom plot.}
    \label{fig:SeparatedResults}
\end{figure}

For $F_\text{obs}(r, \phi)$, we model distance-related fluence decay as $R(r) = R_0\exp(-\alpha r)$, where $r$ is the horizontal distance from the TGF and $\alpha$ is fit empirically.
For $P(\phi)$, we initially use a simple LOESS smoother operating in log space.
We can then iteratively solve the unknowns in each function by setting $R(r) = F_\text{obs}/P(\phi)$ and $P(\phi) = F_\text{obs}/R(r)$.
Once this process converges (which takes six iterations in this case), we replace the LOESS smoother with a model that is the sum of two Gaussians.
The results of this process are shown in \cref{fig:SeparatedResults}, where ``distance-adjusted fluence'' refers to $F_\text{obs}/R(r)$ and ``phi-adjusted fluence'' to $F_\text{obs}/P(\phi)$.
The full model is thus
\begin{align}
  F(r, \phi)_\text{obs} \approx
  R_0 \,
  e^{-\alpha r}
  \left[
    \exp\left(-\frac{(\phi - \phi_1)^2}{2\sigma_1^2}\right) +
    B\exp\left(-\frac{(\phi - \phi_2)^2}{2\sigma_2^2}\right)
  \right].
\label{eq:fullmodel}
\end{align}

\begin{table}
    \caption{Coefficients for the model $F(r, \phi)$}
    \label{tab:parameters}
    \centering
    \renewcommand{\arraystretch}{0.85}
    \begin{tabular}{lccccccc}
        \toprule
        & $R_0$ & $\alpha$ & $\phi_1$ & $\phi_2$ & $\sigma_1$ & $\sigma_2$ & $B$
    \\
        Units & (MeV~/~m$^2$) & (km$^{-1}$) & (deg) & (deg) & (deg) & (deg) & ---
    \\\midrule
        Value & \num{8.35e9} & $1.83$ & $22.3$ & $-97$ & $40$ & $42.5$ & $0.25$
    \\
        Std.~Error $\pm$ & \num{1.9e9} & $0.032$ & $2.59$ & $6.92$ & $2.12$ & $3.92$ & $0.026$
    \\\bottomrule
    \end{tabular}
\end{table}

The estimated numerical values and standard error of the model's parameters are given in \cref{tab:parameters}, and it has a coefficient of determination of $R^2 = 0.946$.
We can compare this model created to fit the data to the results of the \textsc{Geant4} simulations to estimate the following source parameters:

\textbf{Off-nadir angle ($\theta$):}
The best-fit value of $\alpha$ based on data is 1.83, which falls between the simulated values of $\alpha = \qty{1.43}{\kilo\meter}$ for $\theta = \ang{85}$ and $\alpha = \qty{3.07}{\per\kilo\meter}$ for $\theta = \ang{75}$.
A straightforward linear interpolation would suggest $\theta = \ang{82.6}$, though $\alpha = \qty{1.83}{\per\kilo\meter}$ could also be achieved by a linear combination of the simulated $\theta$ values.
A real TGF will obviously have some distribution over $\theta$, but the results of the simulation therefore tell us that this TGF had strong components above $\theta = \ang{75}$.
Since lower $\theta$ angles contribute mostly to the fluence closer than \qty{3}{km}, where detectors were saturated, we cannot constrain low $\theta$ components other than saying they could not have been bright enough to contribute fluence at $r \geq \qty{3}{km}$.

\begin{figure}
    \centering
    \includegraphics[width=\linewidth]{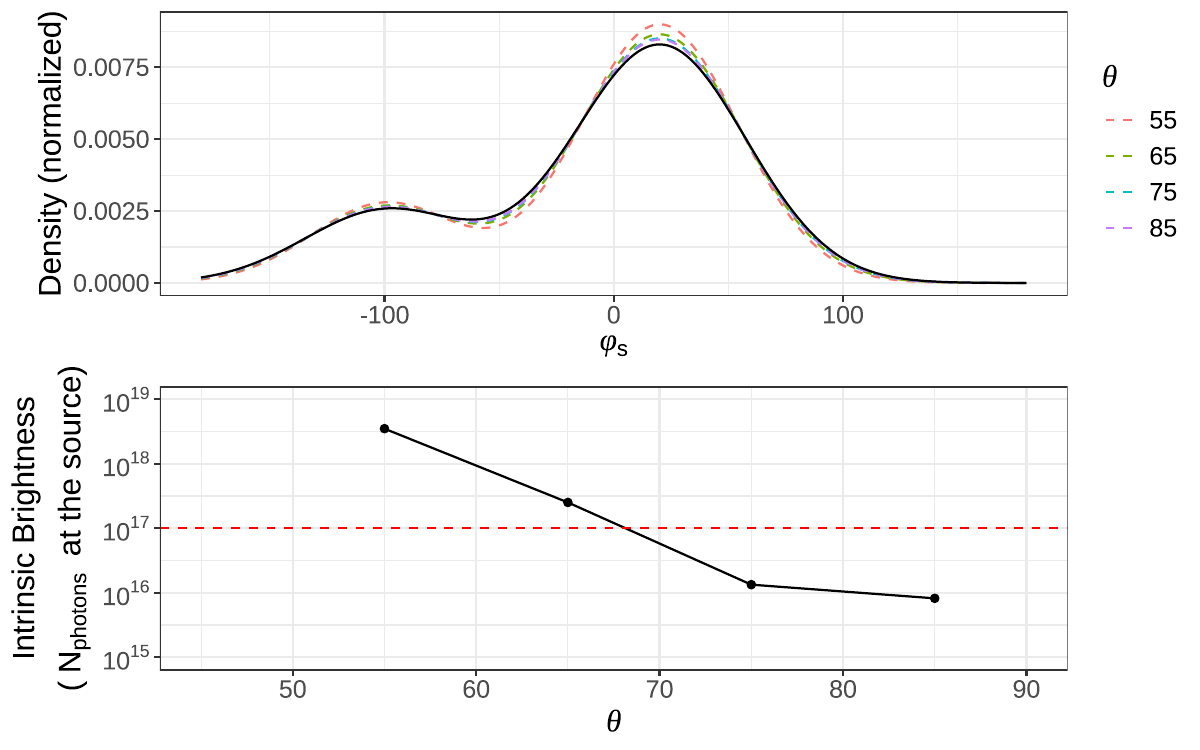}
    \caption{\emph{Top:} Dotted lines show the distribution of photons at the source that, when convolved with the atmospheric response in \textsc{Geant4}, recreate the azimuthal function in black from equation \cref{eq:fullmodel}.
    \emph{Bottom:} Required photons at the source in order to recreate the fluence seen on the ground according to \textsc{Geant4} simulations.
    The dotted red line represents the ``standard'' intrinsic brightness of \num{e17}.}
    \label{fig:sourceParams}
\end{figure}

\textbf{Azimuthal distribution at the source:}
Because we have modeled the $\phi$ dependence of $F_\text{obs}(r, \phi)$ as a smooth function that is the sum of two Gaussians, we can now deconvolve \cref{eq:convolution} to solve for the TGF source distribution in $\phi_\text{s}$ for a given $\theta_i$, $s_{\theta_i}(\phi_\text{s})$.
Deconvolutions are extremely sensitive to noise and would generally fail for a more irregular shaped function for $P(\phi)$.
The deconvolution of two Gaussians, however, is itself a Gaussian, with a slightly smaller $\sigma$ just like we would expect since it should represent the distribution prior to atmospheric scattering.
The deconvolved source distribution is shown for each simulated $\theta_i$ as the dotted lines in the top plot of \cref{fig:sourceParams}.
Those source distributions have increasingly narrow peaks with decreasing $\theta_i$, which is consistent with the fact that we saw an increasingly wide atmospheric response with decreasing $\theta_i$ as in \cref{fig:simulations}.

\textbf{Intrinsic Brightness:}
We model the total energy deposited on the ground over the annulus from 3 to \qty{7}{km} by integrating $F(r, \phi)$ over that region.
To estimate the intrinsic brightness required for each $\theta_i$, we scale the energy deposited in the ground over that region in the simulations to match our observations.
This gives us a lower limit on brightness, for the hypothetical case of a beam tightly centered on $\theta = \ang{85}$, near \num{e16} initial photons above \qty{1}{MeV} as shown in \cref{fig:sourceParams}.
In addition to the previously mentioned disagreements with the slope of $R(r)$, a distribution strongly centered on lower angles $\theta \leq \ang{45}$ would require an unrealistically bright source to explain the calculated energy deposited at the ground.
We cannot confidently put an upper limit on brightness due to nearly all the detectors within \qty{3}{km} of the source saturating from the high flux.

All the evidence, when taken as a whole, tends to point toward the TGF source comprising an emission peak at $\phi_\text{s} = \ang{20}$, and a secondary peak at $\phi_\text{s} = -\ang{100}$ weaker by a factor of 4, where both peaks appear to concentrate at angles $\theta > \ang{75}$ with a brightness of at least \num{e16} initial photons above \qty{1}{MeV}.

\subsection{Physical Motivation for the Model}
\label{sec:theory}

\citet{dwyer07} suggests that if two avalanche regions with opposite electric field vectors were to lie adjacent to each other, the threshold required for relativistic feedback would be lowered significantly.
This is because seed-inducing X-rays would not have to back scatter nor would positrons need to turn around to reach the beginning of another avalanche region.
This idea has been discussed again in the context of a radially directed electric field transverse to the leader channel \citep{KutsykTransverse2011,dwyer12return}, as well as more generally for any adjacent avalanche regions in thunderclouds \citep{stadnichukFeedback21}.

The direction of the avalanche regions in this scenario, largely pointing radially inward toward the lightning channel, has been previously described in \citet{Dwyer2021}.
Although they model a TGF occurring in a field enhanced around a positive leader, the polarity and overall geometry would be the same in the case of a negative return stroke approaching the cloud base.
\citet{dwyer12return} point out that the return stroke would enhance the field greatly by bringing the ground potential up near the main negative in this case, possibly exceeding the runaway threshold.
This alone would account for the enhanced horizontal components of the gamma emission, and the azimuthal asymmetry would follow from any inhomogeneity in the background field and/or tilting of the return stroke channel.
Future studies should explore this geometry in detail and the resulting fluence patterns on the ground, especially as more events from Auger become available.

\section{Conclusion}

In this work, we presented a TGF that was recorded at the Pierre Auger Observatory in Argentina.
The footprint of the TGF was seen by 54 detectors on the ground spanning a diameter of approximately \qty{14}{km}, whereas most TGFs in the literature have only one observation point.
The high degree of spatial coverage allowed us to do two unique analyses.
First, we were able to use the time of arrival of the gamma rays to triangulate the source position of the TGF itself, located at \qty{2.3\pm0.5}{km} above ground level.
Second, we were able to analyze the fluence distribution relative to the TGF source to identify an azimuthal asymmetry with two distinct peaks at $\phi = \ang{-20}$ and $\phi = \ang{100}$.
Herein lies the most surprising result: not that the TGF was asymmetric about is vertical axis with the ground, but that there appears to be no axis about which the TGF emission could be azimuthally symmetric at all.

We used Geant4 simulations to come up with a range of angular distributions at the source that could be consistent with the fluence pattern seen on the ground.
The simulations suggest that the TGF had a large component of its angular distribution at $\theta > \ang{75}$, with a lower limit on brightness for the TGF of ${\geq}10^{16}$ gamma rays above \qty{1}{MeV}.
By treating the fluence on the ground as a convolution of the source distribution and the simulated atmospheric response, we were also able to estimate how much narrower the azimuthal distribution and that TGF source was than the distribution seen on the ground. 

Given the scarcity of multipoint TGF measurements it is impossible to say whether or not fluence patterns like these are common.
We suspect this asymmetric, high-angle emission would be more common in TGFs induced by a return stroke as mentioned in \cref{sec:theory}.
This would help to explain the ``laterally distant'' observations in \citet{Ortberg23}; the authors found that even an isotropic TGF at standard brightness was far too weak to be seen by their detectors.
Though at Utah's Telescope Array they tend to see smaller footprints and have not done formal analysis on angular patterns, some of their published events appear to be irregularly shaped \citep{abbasi18, 2020BelzUtahTAB}.
The +IC induced TGFs commonly seen from space likely differ in source geometry -- the environment is thought to be more strongly dominated by a consistent background field with relatively less extreme enhancement by a propagating stepped leader vs.\ a return stroke.
Still, the extent of any axial asymmetry of such events remains unknown.
In addition to the continued serendipity of coincident detections between different satellites, several satellite arrays capable of multipoint detection have been proposed \citep{2019PanasyukSOCRAT, 2021Tryad}.
These multipoint detections are the most effective, if not the only, way to learn about the source geometry of the avalanche region(s) of the observed TGF.

\acknowledgments

This work was supported by a grant from the National Science Foundation, ATM-0607885.

\begin{sloppypar}
The successful installation, commissioning, and operation of the Pierre
Auger Observatory would not have been possible without the strong
commitment and effort from the technical and administrative staff in
Malarg\"ue. We are very grateful to the following agencies and
organizations for financial support:
\end{sloppypar}

\begin{sloppypar}
Argentina -- Comisi\'on Nacional de Energ\'\i{}a At\'omica; Agencia Nacional de
Promoci\'on Cient\'\i{}fica y Tecnol\'ogica (ANPCyT); Consejo Nacional de
Investigaciones Cient\'\i{}ficas y T\'ecnicas (CONICET); Gobierno de la
Provincia de Mendoza; Municipalidad de Malarg\"ue; NDM Holdings and Valle
Las Le\~nas; in gratitude for their continuing cooperation over land
access; Australia -- the Australian Research Council; Belgium -- Fonds
de la Recherche Scientifique (FNRS); Research Foundation Flanders (FWO),
Marie Curie Action of the European Union Grant No.~101107047; Brazil --
Minist\'erio da Ci\^encia, Tecnologia e Inova\c{c}\~ao (MCTI); Czech Republic --
GACR 24-13049S, CAS LQ100102401, MEYS LM2023032,
CZ.02.1.01/0.0/0.0/16{\textunderscore}013/0001402, CZ.02.1.01/0.0/0.0/18{\textunderscore}046/0016010
and CZ.02.1.01/0.0/0.0/17{\textunderscore}049/0008422 and
CZ.02.01.01/00/22{\textunderscore}008/0004632; France -- Centre de Calcul IN2P3/CNRS;
Centre National de la Recherche Scientifique (CNRS); Institut National
de Physique Nucl\'eaire et de Physique des Particules (IN2P3/CNRS);
Germany -- Bundesministerium f\"ur Forschung, Technologie und Raumfahrt
(BMFTR); Deutsche Forschungsgemeinschaft (DFG); Ministerium f\"ur Finanzen
Baden-W\"urttemberg; Helmholtz Alliance for Astroparticle Physics (HAP);
Hermann von Helmholtz-Gemeinschaft Deutscher Forschungszentren e.V.;
Ministerium f\"ur Kultur und Wissenschaft des Landes Nordrhein-Westfalen;
Ministerium f\"ur Wissenschaft, Forschung und Kunst des Landes
Baden-W\"urttemberg; Italy -- Istituto Nazionale di Fisica Nucleare
(INFN); Istituto Nazionale di Astrofisica (INAF); Ministero
dell'Universit\`a e della Ricerca (MUR); CETEMPS Center of Excellence;
Ministero degli Affari Esteri (MAE), ICSC Centro Nazionale di Ricerca in
High Performance Computing, Big Data and Quantum Computing, funded by
European Union NextGenerationEU, reference code CN{\textunderscore}00000013; M\'exico --
Consejo Nacional de Ciencia y Tecnolog\'\i{}a (CONACYT-SECHTI)
No.~CB-A1-S-46703, Universidad Nacional Aut\'onoma de M\'exico (UNAM)
PAPIIT-IN114924; Benem\'erita Universidad Aut\'onoma de Puebla (BUAP), VIEP
and Laboratorio Nacional de Superc\'omputo del Sureste de M\'exico (LNS);
and Benem\'erita Universidad Aut\'onoma de Chiapas (UNACH); The Netherlands
-- Ministry of Education, Culture and Science; Netherlands Organisation
for Scientific Research (NWO); Dutch national e-infrastructure with the
support of SURF Cooperative; Poland -- Ministry of Science and Higher
Education, grant No.~2022/WK/12; National Science Centre, grants
No.~2020/39/B/ST9/01398, and 2022/45/B/ST9/02163; Portugal -- Portuguese
national funds and FEDER funds within Programa Operacional Factores de
Competitividade through Funda\c{c}\~ao para a Ci\^encia e a Tecnologia
(COMPETE); Romania -- Ministry of Education and Research, contract
no.~30N/2023 under Romanian National Core Program LAPLAS VII, and grant
no.~PN 23 21 01 02; Slovenia -- Slovenian Research and Innovation
Agency, grants P1-0031, I0-0033; Spain -- Ministerio de Ciencia,
Innovaci\'on y Universidades/Agencia Estatal de Investigaci\'on MICIU/AEI
/10.13039/501100011033 (PID2022-140510NB-I00, PCI2023-145952-2,
CNS2024-154676, and Mar\'\i{}a de Maeztu CEX2023-001318-M), Xunta de Galicia
(CIGUS Network of Research Centers, Consolidaci\'on ED431C-2025/11 and
ED431F-2022/15) and European Union ERDF; USA -- Department of Energy,
Contracts No.~DE-AC02-07CH11359, No.~DE-FR02-04ER41300,
No.~DE-FG02-99ER41107 and No.~DE-SC0011689; National Science Foundation,
Grant No.~0450696, and NSF-2013199; The Grainger Foundation;
Astrophysics Centre for Multi-messenger studies in Europe (ACME) EU
Grant No 101131928; and UNESCO.
\end{sloppypar}

\noindent 
\textbf{Open Research}

Data and materials associated with this work are available at \url{https://doi.org/10.5281/zenodo.21079890} \citep{Ortberg2026Zenodo}.

\bibliography{tgfref.bib}

@string{grl = "Geophys. Res. Let."}

@string{science = "Science"}

@string{prl = "Phys. Rev. Lett."}

@string{nat = "Nature"}

@INPROCEEDINGS{2021Tryad,
       author = {{Lupo}, Alexis and {Jenke}, Peter and {Briggs}, Michael},
        title = "{Building Gamma-Ray Detectors for Two CubeSats}",
    booktitle = {AGU Fall Meeting Abstracts},
         year = 2021,
       volume = {2021},
        month = dec,
          eid = {AE15A-1886},
        pages = {AE15A-1886},
       adsurl = {https://ui.adsabs.harvard.edu/abs/2021AGUFMAE15A1886L}
}

@ARTICLE{2019PanasyukSOCRAT,
       author = {{Panasyuk}, Mikhail and {Klimov}, Pavel and {Svertilov}, Sergei and {Belov}, Alexander and {Bogomolov}, Vitali and {Bogomolov}, Andrei and {Garipov}, Gali and {Iyudin}, Anatoly and {Kaznacheeva}, Margarita and {Maksimov}, Ivan and {Minaev}, Alexander and {Novikov}, Artem and {Minaev}, Pavel and {Petrov}, Vasili and {Pozanenko}, Alexei and {Shtunder}, Yan and {Yashin}, Ivan},
        title = "{Universat-SOCRAT multi-satellite project to study TLEs and TGFs}",
      journal = {Progress in Earth and Planetary Science},
         year = 2019,
        month = dec,
       volume = {6},
       number = {1},
          eid = {35},
        pages = {35},
          doi = {10.1186/s40645-019-0280-3},
       adsurl = {https://ui.adsabs.harvard.edu/abs/2019PEPS....6...35P}
}

@ARTICLE{KutsykTransverse2011,
       author = {{Kutsyk}, I.~M. and {Babich}, L.~P. and {Donskoi}, E.~N.},
        title = "{Self-sustained relativistic-runaway-electron avalanches in the transverse field of lightning leader as sources of terrestrial gamma-ray flashes}",
      journal = {Soviet Journal of Experimental and Theoretical Physics Letters},
         year = 2011,
        month = dec,
       volume = {94},
       number = {8},
        pages = {606-609},
          doi = {10.1134/S0021364011200094},
       adsurl = {https://ui.adsabs.harvard.edu/abs/2011JETPL..94..606K}
}

@inproceedings{Marisaldi20,
       author = {{Marisaldi}, M. and {Mezentsev}, A. and {Sarria}, Sr., D. and {Lindanger}, A. and {Ostgaard}, N. and {Neubert}, T. and {Victor}, R. and {Kochkin}, P. and {Lehtinen}, N.~G. and {Maiorana}, C. and {Skeie}, C.~A.~A. and {Bj{\o}rge-Engeland}, I. and {Ullaland}, K. and {Genov}, G. and {Christiansen}, F. and {Christian}, Jr., H.~J. and {Alnussirat}, S. and {Briggs}, M.~S. and {Ursi}, A. and {Tavani}, M.},
        title = "{ASIM - Fermi - AGILE simultaneous observation of Terrestrial Gamma-ray Flashes}",
    booktitle = {AGU Fall Meeting Abstracts},
         year = 2020,
       series = {AGU Fall Meeting Abstracts},
       volume = {2020},
        month = dec,
          eid = {AE005-04},
        pages = {AE005-04},
       adsurl = {https://ui.adsabs.harvard.edu/abs/2020AGUFMAE005..04M}
}

@article{Dwyer2021,
  title = {Terrestrial gamma-ray flashes initiated by positive leaders},
  author = {Dwyer, Joseph R.},
  journal = {Phys. Rev. D},
  volume = {104},
  issue = {4},
  pages = {043012},
  numpages = {10},
  year = {2021},
  month = {Aug},
  publisher = {American Physical Society},
  doi = {10.1103/PhysRevD.104.043012},
  url = {https://link.aps.org/doi/10.1103/PhysRevD.104.043012}
}

@dataset{Ortberg2026Zenodo,
  author       = {Ortberg, John},
  title        = {A Terrestrial Gamma-ray Flash with Multiple Source Regions at the Pierre Auger Observatory},
  month        = jun,
  year         = 2026,
  publisher    = {Zenodo},
  doi          = {10.5281/zenodo.21079890},
  url          = {https://doi.org/10.5281/zenodo.21079890}
}

@article{Zhang2021NbeTgf,
author = {Zhang, Hongbo and Lu, Gaopeng and Lyu, Fanchao and Xiong, Shaolin and Ahmad, Mohd Riduan and Yi, Qibin and Li, Dongshuai and Xian, Tao and Yang, Jing and Liu, Feifan and Zhu, Baoyou and Pu, Yunjiao and Cummer, Steven A. and Briggs, Michael S. and Qie, Xiushu},
title = {On the Terrestrial Gamma-Ray Flashes Preceding Narrow Bipolar Events},
journal = {Geophysical Research Letters},
volume = {48},
number = {8},
pages = {e2020GL092160},
year = {2021},
doi = {https://doi.org/10.1029/2020GL092160},
url = {https://agupubs.onlinelibrary.wiley.com/doi/abs/10.1029/2020GL092160},
eprint = {https://agupubs.onlinelibrary.wiley.com/doi/pdf/10.1029/2020GL092160},
note = {e2020GL092160 2020GL092160}
}

@article{Berge2019DownwardSims,
author = {Berge, N. and Celestin, S.},
title = {Constraining Downward Terrestrial Gamma Ray Flashes Using Ground-Based Particle Detector Arrays},
journal = {Geophysical Research Letters},
volume = {46},
number = {14},
pages = {8424-8430},
doi = {https://doi.org/10.1029/2019GL083252},
url = {https://agupubs.onlinelibrary.wiley.com/doi/abs/10.1029/2019GL083252},
eprint = {https://agupubs.onlinelibrary.wiley.com/doi/pdf/10.1029/2019GL083252},
year = {2019}
}

@article{Mailyan2019,
author = {Mailyan, B. G. and Xu, W. and Celestin, S. and Briggs, M. S. and Dwyer, J. R. and Cramer, E. S. and Roberts, O. J. and Stanbro, M.},
title = {Analysis of Individual Terrestrial Gamma-Ray Flashes With Lightning Leader Models and Fermi Gamma-Ray Burst Monitor Data},
journal = {Journal of Geophysical Research: Space Physics},
volume = {124},
number = {8},
pages = {7170-7183},
doi = {https://doi.org/10.1029/2019JA026912},
url = {https://agupubs.onlinelibrary.wiley.com/doi/abs/10.1029/2019JA026912},
eprint = {https://agupubs.onlinelibrary.wiley.com/doi/pdf/10.1029/2019JA026912},
year = {2019}
}

@INPROCEEDINGS{OstgardALOFT2023,
       author = {{Ostgaard}, Nikolai and {Marisaldi}, Martino and {Ullaland}, Kjetil and {Yang}, Shiming and {Hasan Qureshi}, Bilal and {S{\o}ndergaard}, Jens and {Mezentsev}, Andrey and {Sarria}, David and {Lehtinen}, Nikolai and {Lang}, Timothy and {Christian}, Hugh and {Quick}, Mason and {Blakeslee}, Richard and {Grove}, J. Eric and {Shy}, Daniel},
        title = "{The ALOFT mission: a flight campaign for TGF and gamma-ray glow observations over Central America and the Caribbean in July 2023}",
    booktitle = {EGU General Assembly Conference Abstracts},
         year = 2023,
       series = {EGU General Assembly Conference Abstracts},
        month = may,
          eid = {EGU-3116},
        pages = {EGU-3116},
          doi = {10.5194/egusphere-egu23-3116},
       adsurl = {https://ui.adsabs.harvard.edu/abs/2023EGUGA..25.3116O}
}

@article{fabro19,
author = {Fabró, Ferran and Montanyà, Joan and van der Velde, Oscar A. and Pineda, Nicolau and Williams, Earle R.},
title = {On the TGF/Lightning Ratio Asymmetry},
journal = {Journal of Geophysical Research: Atmospheres},
volume = {124},
number = {12},
pages = {6518-6531},
doi = {https://doi.org/10.1029/2018JD030075},
url = {https://agupubs.onlinelibrary.wiley.com/doi/abs/10.1029/2018JD030075},
eprint = {https://agupubs.onlinelibrary.wiley.com/doi/pdf/10.1029/2018JD030075},
year = {2019}
}

@article{briggs2013,
  author = {Briggs, Michael S. and Xiong, Shaolin and Connaughton, Valerie and Tierney, Dave and Fitzpatrick, Gerard and Foley, Suzanne and Grove, J. Eric and Chekhtman, Alexandre and Gibby, Melissa and Fishman, Gerald J. and McBreen, Shelia and Chaplin, Vandiver L. and Guiriec, Sylvain and Layden, Emily and Bhat, P. N. and Hughes, Maximilian and Greiner, Jochen and Kienlin, Andreas von and Kippen, R. Marc and Meegan, Charles A. and Paciesas, William S. and Preece, Robert D. and Wilson‐Hodge, Colleen and Holzworth, Robert H. and Hutchins, Michael L.},
  year = {2013},
  title = {Terrestrial gamma‐ray flashes in the Fermi era: Improved observations and analysis methods},
  journal = {Journal of Geophysical Research: Space Physics},
  publisher = {Wiley Online Library},
  issn = {2169-9402},
  doi = {10.1002/jgra.50205},
  volume = {118},
  month = {6},
  pages = {3805--3830},
  number = {6},
  url = {https://doi.org/10.1002/jgra.50205},
}

@article{stadnichukFeedback21,
author = {Stadnichuk, E. and Svechnikova, E. and Nozik, A. and Zemlianskaya, D. and Khamitov, T. and Zelenyy, M. and Dolgonosov, M.},
title = {Relativistic Runaway Electron Avalanches Within Complex Thunderstorm Electric Field Structures},
journal = {Journal of Geophysical Research: Atmospheres},
volume = {126},
number = {24},
pages = {e2021JD035278},
doi = {https://doi.org/10.1029/2021JD035278},
url = {https://agupubs.onlinelibrary.wiley.com/doi/abs/10.1029/2021JD035278},
eprint = {https://agupubs.onlinelibrary.wiley.com/doi/pdf/10.1029/2021JD035278},
note = {e2021JD035278 2021JD035278},
year = {2021}
}

@ARTICLE{mailyan16,
   author = {{Mailyan}, B.~G. and {Briggs}, M.~S. and {Cramer}, E.~S. and 
	{Fitzpatrick}, G. and {Roberts}, O.~J. and {Stanbro}, M. and 
	{Connaughton}, V. and {McBreen}, S. and {Bhat}, P.~N. and {Dwyer}, J.~R.
	},
    title = "{The spectroscopy of individual terrestrial gamma-ray flashes: Constraining the source properties}",
  journal = {Journal of Geophysical Research (Space Physics)},
     year = 2016,
    month = nov,
   volume = 121,
   number = {A10},
    pages = {11},
      doi = {10.1002/2016JA022702},
   adsurl = {http://adsabs.harvard.edu/abs/2016JGRA..12111346M}
}

@article{lindanger21,
author = {Lindanger, A. and Marisaldi, M. and Sarria, D. and Østgaard, N. and Lehtinen, N. and Skeie, C. A. and Mezentzev, A. and Kochkin, P. and Ullaland, K. and Yang, S. and Genov, G. and Carlson, B. E. and Köhn, C. and Navarro-Gonzalez, J. and Connell, P. and Reglero, V. and Neubert, T.},
title = {Spectral Analysis of Individual Terrestrial Gamma-Ray Flashes Detected by ASIM},
journal = {Journal of Geophysical Research: Atmospheres},
volume = {126},
number = {23},
pages = {e2021JD035347},
doi = {https://doi.org/10.1029/2021JD035347},
url = {https://agupubs.onlinelibrary.wiley.com/doi/abs/10.1029/2021JD035347},
eprint = {https://agupubs.onlinelibrary.wiley.com/doi/pdf/10.1029/2021JD035347},
note = {e2021JD035347 2021JD035347},
year = {2021}
}

@article{hazelton09,
	author ={B.J. Hazelton and B.W. Grefenstette and D.M. Smith and J.R. Dwyer and
                X.-M. Shao and S.A. Cummer and T. Chronis and E.H. Lay and R.H. Holzworth},
	title =	{Spectral dependence of terrestrial gamma-ray flashes on source distance},
	journal = grl,
        volume={36},
	year = {2009},
	pages = {L01108},
        doi={10.1029/2008GL035906}
}

@ARTICLE{gjest11,
   author = {{Gjesteland}, T. and {{\O}stgaard}, N. and {Collier}, A.~B. and 
	{Carlson}, B.~E. and {Cohen}, M.~B. and {Lehtinen}, N.~G.},
    title = "{Confining the angular distribution of terrestrial gamma ray flash emission}",
  journal = {Journal of Geophysical Research (Space Physics)},
     year = 2011,
    month = nov,
   volume = 116,
      eid = {A11313},
    pages = {11313},
      doi = {10.1029/2011JA016716},
   adsurl = {http://adsabs.harvard.edu/abs/2011JGRA..11611313G}
}

@article{Albrechtsen2018,
   author = {{Albrechtsen}, K.~H. and {{\O}stgaard}, N. and {Berge}, N. and 
	{Gjesteland}, T.},
    title = "{Observationally Weak TGFs in the RHESSI Data}",
  journal = {Journal of Geophysical Research (Atmospheres)},
     year = 2019,
    month = jan,
   volume = 124,
    pages = {287-298},
      doi = {10.1029/2018JD029272},
   adsurl = {http://adsabs.harvard.edu/abs/2019JGRD..124..287A}
}

@article{dwyersmith05,
	author =	{J. R. Dwyer and D. M. Smith},
	title =	{A Comparison between {M}onte {C}arlo simulations of runaway breakdown
			and terrestrial gamma-ray flash observations},
	journal = grl,
	volume =	{32},
	doi =		{10.1028/2005GL023848},
	pages =		{L08811},
	year =	{2005}
}

@article{ostgaard19,
author = {{\O}stgaard, N. and Neubert, T. and Reglero, V. and Ullaland, K. and Yang, S. and Genov, G. and Marisaldi, M. and Mezentsev, A. and Kochkin, P. and Lehtinen, N. and Sarria, D. and Qureshi, B. H. and Solberg, A. and Maiorana, C. and Albrechtsen, K. and Budtz-J{\o}rgensen, C. and Kuvvetli, I. and Christiansen, F. and Chanrion, O. and Heumesser, M. and Navarro-Gonzalez, J. and Connell, P. and Eyles, C. and Christian, H. and Al-nussirat, S.},
title = {First 10 Months of TGF Observations by ASIM},
journal = {Journal of Geophysical Research: Atmospheres},
pages = {},
year = {2019},
doi = {10.1029/2019JD031214},
url = {https://agupubs.onlinelibrary.wiley.com/doi/abs/10.1029/2019JD031214},
eprint = {https://agupubs.onlinelibrary.wiley.com/doi/pdf/10.1029/2019JD031214}
}

@ARTICLE{enoto17,
   author = {{Enoto}, T. and {Wada}, Y. and {Furuta}, Y. and {Nakazawa}, K. and 
	{Yuasa}, T. and {Okuda}, K. and {Makishima}, K. and {Sato}, M. and 
	{Sato}, Y. and {Nakano}, T. and {Umemoto}, D. and {Tsuchiya}, H.
	},
    title = "{Photonuclear reactions triggered by lightning discharge}",
  journal = nat,
     year = 2017,
    month = nov,
   volume = 551,
    pages = {481-484},
      doi = {10.1038/nature24630},
   adsurl = {http://adsabs.harvard.edu/abs/2017Natur.551..481E}
}

@ARTICLE{bowers17,
   author = {{Bowers}, G.~S. and {Smith}, D.~M. and {Martinez-McKinney}, G.~F. and 
	{Kamogawa}, M. and {Cummer}, S.~A. and {Dwyer}, J.~R. and {Wang}, D. and 
	{Stock}, M. and {Kawasaki}, Z.},
    title = "{Gamma Ray Signatures of Neutrons From a Terrestrial Gamma Ray Flash}",
  journal = grl,
     year = 2017,
    month = oct,
   volume = 44,
    pages = {10},
      doi = {10.1002/2017GL075071},
   adsurl = {http://adsabs.harvard.edu/abs/2017GeoRL..4410063B}
}

@article{Smith18,
author = {Smith, D. and Bowers, Gregory and Kamogawa, Masashi and Wang, D. and Ushio, T. and Ortberg, John and Dwyer, J. and Stock, M.},
year = {2018},
month = {09},
pages = {},
title = {Characterizing Upward Lightning With and Without a Terrestrial Gamma Ray Flash},
volume = {123},
journal = {Journal of Geophysical Research: Atmospheres},
doi = {10.1029/2018JD029105}
}

@ARTICLE{hare16,
   author = {{Hare}, B.~M. and {Uman}, M.~A. and {Dwyer}, J.~R. and {Jordan}, D.~M. and 
	{Biggerstaff}, M.~I. and {Caicedo}, J.~A. and {Carvalho}, F.~L. and 
	{Wilkes}, R.~A. and {Kotovsky}, D.~A. and {Gamerota}, W.~R. and 
	{Pilkey}, J.~T. and {Ngin}, T.~K. and {Moore}, R.~C. and {Rassoul}, H.~K. and 
	{Cummer}, S.~A. and {Grove}, J.~E. and {Nag}, A. and {Betten}, D.~P. and 
	{Bozarth}, A.},
    title = "{Ground-level observation of a terrestrial gamma ray flash initiated by a triggered lightning}",
  journal = {Journal of Geophysical Research (Atmospheres)},
     year = 2016,
    month = jun,
   volume = 121,
    pages = {6511-6533},
      doi = {10.1002/2015JD024426},
   adsurl = {http://adsabs.harvard.edu/abs/2016JGRD..121.6511H}
}

@ARTICLE{wada18,
   author = {{Wada}, Y. and {Bowers}, G.~S. and {Enoto}, T. and {Kamogawa}, M. and 
	{Nakamura}, Y. and {Morimoto}, T. and {Smith}, D.~M. and {Furuta}, Y. and 
	{Nakazawa}, K. and {Yuasa}, T. and {Matsuki}, A. and {Kubo}, M. and 
	{Tamagawa}, T. and {Makishima}, K. and {Tsuchiya}, H.},
    title = "{Termination of Electron Acceleration in Thundercloud by Intra/Inter-cloud Discharge}",
  journal = {ArXiv e-prints},
archivePrefix = "arXiv",
   eprint = {1805.04721},
 primaryClass = "physics.ao-ph",
     year = 2018,
    month = may,
   adsurl = {http://adsabs.harvard.edu/abs/2018arXiv180504721W}
}

@ARTICLE{wada19,
       author = {{Wada}, Y. and {Enoto}, T. and {Nakazawa}, K. and {Furuta}, Y. and
         {Yuasa}, T. and {Nakamura}, Y. and {Morimoto}, T. and {Matsumoto}, T. and
         {Makishima}, K. and {Tsuchiya}, H.},
        title = "{Downward Terrestrial Gamma-Ray Flash Observed in a Winter Thunderstorm}",
      journal = prl,
         year = "2019",
        month = "Aug",
       volume = {123},
       number = {6},
          doi = {10.1103/PhysRevLett.123.061103},
archivePrefix = {arXiv},
       eprint = {1907.06239},
 primaryClass = {physics.ao-ph},
       adsurl = {https://ui.adsabs.harvard.edu/abs/2019PhRvL.123f1103W}
}

@ARTICLE{dwyer12return,
   author = {{Dwyer}, J.~R. and {Schaal}, M.~M. and {Cramer}, E. and {Arabshahi}, S. and 
	{Liu}, N. and {Rassoul}, H.~K. and {Hill}, J.~D. and {Jordan}, D.~M. and 
	{Uman}, M.~A.},
    title = "{Observation of a gamma-ray flash at ground level in association with a cloud-to-ground lightning return stroke}",
  journal = {Journal of Geophysical Research (Space Physics)},
     year = 2012,
    month = oct,
   volume = 117,
      eid = {A10303},
    pages = {10303},
      doi = {10.1029/2012JA017810},
   adsurl = {http://adsabs.harvard.edu/abs/2012JGRA..11710303D}
}

@article{abbasi18,
author={{Abbasi}, R.~U. and others },
title="{Gamma‐ray Showers Observed at Ground Level in Coincidence With Downward Lightning Leaders}",
journal={Journal of Geophysical Research (Atmospheres)},
year=2018,
doi={10.1029/2017JD027931},
}

@ARTICLE{dwyer12tgftheory,
   author = {{Dwyer}, J.~R.},
    title = "{The relativistic feedback discharge model of terrestrial gamma ray flashes}",
  journal = {Journal of Geophysical Research (Space Physics)},
     year = 2012,
    month = feb,
   volume = 117,
      eid = {A02308},
    pages = {2308},
      doi = {10.1029/2011JA017160},
   adsurl = {http://adsabs.harvard.edu/abs/2012JGRA..117.2308D}
}

@ARTICLE{celestin11,
   author = {{Celestin}, S. and {Pasko}, V.~P.},
    title = "{Energy and fluxes of thermal runaway electrons produced by exponential growth of streamers during the stepping of lightning leaders and in transient luminous events}",
  journal = {Journal of Geophysical Research (Space Physics)},
     year = 2011,
    month = mar,
   volume = 116,
      eid = {A03315},
    pages = {3315},
      doi = {10.1029/2010JA016260},
   adsurl = {http://adsabs.harvard.edu/abs/2011JGRA..116.3315C}
}

@article{dwyer07,
	author = {J. R. Dwyer}, 
	title = {Relativistic breakdown in planetary atmospheres},
	journal = {Physics of Plasmas},
	volume = {14},
	pages = {042901}, 
	doi = {10.1063/1.2709652},
	year = {2007}
}

@article{2017Abbasi,
title = {The bursts of high energy events observed by the telescope array surface detector},
journal = {Physics Letters A},
volume = {381},
number = {32},
pages = {2565-2572},
year = {2017},
issn = {0375-9601},
doi = {https://doi.org/10.1016/j.physleta.2017.06.022},
url = {https://www.sciencedirect.com/science/article/pii/S0375960117305893},
author = {R.U. Abbasi and M. Abe and T. Abu-Zayyad and M. Allen and R. Anderson and R. Azuma and E. Barcikowski and J.W. Belz and D.R. Bergman and S.A. Blake and R. Cady and B.G. Cheon and J. Chiba and M. Chikawa and T. Fujii and M. Fukushima and T. Goto and W. Hanlon and Y. Hayashi and N. Hayashida and K. Hibino and K. Honda and D. Ikeda and N. Inoue and T. Ishii and R. Ishimori and H. Ito and D. Ivanov and C.C.H. Jui and K. Kadota and F. Kakimoto and O. Kalashev and K. Kasahara and H. Kawai and S. Kawakami and S. Kawana and K. Kawata and E. Kido and H.B. Kim and J.H. Kim and J.H. Kim and S. Kishigami and S. Kitamura and Y. Kitamura and V. Kuzmin and Y.J. Kwon and J. Lan and J.P. Lundquist and K. Machida and K. Martens and T. Matsuda and T. Matsuyama and J.N. Matthews and M. Minamino and K. Mukai and I. Myers and K. Nagasawa and S. Nagataki and T. Nakamura and T. Nonaka and A. Nozato and S. Ogio and J. Ogura and M. Ohnishi and H. Ohoka and K. Oki and T. Okuda and M. Ono and R. Onogi and A. Oshima and S. Ozawa and I.H. Park and M.S. Pshirkov and D.C. Rodriguez and G. Rubtsov and D. Ryu and H. Sagawa and K. Saito and Y. Saito and N. Sakaki and N. Sakurai and A.L. Sampson and L.M. Scott and K. Sekino and P.D. Shah and F. Shibata and T. Shibata and H. Shimodaira and B.K. Shin and H.S. Shin and J.D. Smith and P. Sokolsky and R.W. Springer and B.T. Stokes and S.R. Stratton and T.A. Stroman and T. Suzawa and M. Takamura and M. Takeda and R. Takeishi and A. Taketa and M. Takita and Y. Tameda and H. Tanaka and K. Tanaka and M. Tanaka and S.B. Thomas and G.B. Thomson and P. Tinyakov and I. Tkachev and H. Tokuno and T. Tomida and S. Troitsky and Y. Tsunesada and K. Tsutsumi and Y. Uchihori and S. Udo and F. Urban and G. Vasiloff and T. Wong and R. Yamane and H. Yamaoka and K. Yamazaki and J. Yang and K. Yashiro and Y. Yoneda and S. Yoshida and H. Yoshii and R. Zollinger and Z. Zundel}
}

@article{AugerBackground2015,
author = {{Pierre Auger Collaboration}},
title = {The Pierre Auger Cosmic Ray Observatory},
journal = {Nuclear Instruments and Methods in Physics Research Section A: Accelerators, Spectrometers, Detectors and Associated Equipment},
volume = {798},
pages = {172-213},
year = {2015},
issn = {0168-9002},
doi = {https://doi.org/10.1016/j.nima.2015.06.058},
url = {https://www.sciencedirect.com/science/article/pii/S0168900215008086}
}

@article{2006SDCalibration,
author = {{Pierre Auger Collaboration}},
title = {The Pierre Auger Cosmic Ray Observatory},
journal = {Nuclear Instruments and Methods in Physics Research Section A: Accelerators, Spectrometers, Detectors and Associated Equipment},
volume = {568},
pages = {839},
year = {2006}
}

@INPROCEEDINGS{2017AugerTGFsColalillo,
       author = {{Colalillo}, R. and {Pierre Auger Collaboration}},
        title = "{Peculiar lightning-related events observed by the surface detector of the Pierre Auger Observatory}",
    booktitle = {35th International Cosmic Ray Conference (ICRC2017)},
         year = 2017,
       series = {International Cosmic Ray Conference},
       volume = {301},
        month = jul,
          eid = {314},
        pages = {314},
          doi = {10.22323/1.301.0314},
       adsurl = {https://ui.adsabs.harvard.edu/abs/2017ICRC...35..314C}
}

@INPROCEEDINGS{AugerAtmoElectricity2023,
       author = {{Colalillo}, R. and {Pierre Auger Collaboration}},
        title = "{The Pierre Auger Observatory: Studying atmospheric electricity with cosmic-ray detectors}",
    booktitle = {European Physical Journal Web of Conferences},
         year = 2023,
       series = {European Physical Journal Web of Conferences},
       volume = {283},
        month = oct,
          eid = {06014},
        pages = {06014},
          doi = {10.1051/epjconf/202328306014},
       adsurl = {https://ui.adsabs.harvard.edu/abs/2023EPJWC.28306014C}
}

@ARTICLE{AugerElves2020,
       author = {{Pierre Auger Collaboration}},
        title = "{A 3-Year Sample of Almost 1,600 Elves Recorded Above South America by the Pierre Auger Cosmic-Ray Observatory}",
      journal = {Earth and Space Science},
         year = 2020,
        month = apr,
       volume = {7},
       number = {4},
          eid = {e00582},
        pages = {e00582},
          doi = {10.1029/2019EA000582},
       adsurl = {https://ui.adsabs.harvard.edu/abs/2020E&SS....700582A}
}

@article{Schimassek_2022,
doi = {10.1088/1742-6596/2398/1/012003},
url = {https://dx.doi.org/10.1088/1742-6596/2398/1/012003},
year = {2022},
month = {dec},
publisher = {IOP Publishing},
volume = {2398},
number = {1},
pages = {012003},
author = {{Schimassek}, M. and {Pierre Auger Collaboration}},
title = {Terrestrial Gamma-Ray Flashes at the Pierre Auger Observatory},
journal = {Journal of Physics: Conference Series}
}

@article{Chaffin2024,
author = {Chaffin, Jeffrey M. and Smith, David M. and Lapierre, Jeff and Cummer, Steve and Ortberg, John and Sunjerga, Antonio and Mostajabi, Amirhossein and Rubinstein, Marcos and Rachidi, Farhad},
title = {Mountaintop Gamma Ray Observations of Three Terrestrial Gamma-Ray Flashes at the Säntis Tower, Switzerland With Coincident Radio Waveforms},
journal = {Journal of Geophysical Research: Atmospheres},
volume = {129},
number = {2},
pages = {e2023JD039761},
doi = {https://doi.org/10.1029/2023JD039761},
url = {https://agupubs.onlinelibrary.wiley.com/doi/abs/10.1029/2023JD039761},
eprint = {https://agupubs.onlinelibrary.wiley.com/doi/pdf/10.1029/2023JD039761},
note = {e2023JD039761 2023JD039761},
year = {2024}
}

@article{Wada2020,
author = {Wada, Y. and Enoto, T. and Nakamura, Y. and Morimoto, T. and Sato, M. and Ushio, T. and Nakazawa, K. and Yuasa, T. and Yonetoku, D. and Sawano, T. and Kamogawa, M. and Sakai, H. and Furuta, Y. and Makishima, K. and Tsuchiya, H.},
title = {High Peak-Current Lightning Discharges Associated With Downward Terrestrial Gamma-Ray Flashes},
journal = {Journal of Geophysical Research: Atmospheres},
volume = {125},
number = {4},
pages = {e2019JD031730},
doi = {https://doi.org/10.1029/2019JD031730},
url = {https://agupubs.onlinelibrary.wiley.com/doi/abs/10.1029/2019JD031730},
eprint = {https://agupubs.onlinelibrary.wiley.com/doi/pdf/10.1029/2019JD031730},
note = {e2019JD031730 10.1029/2019JD031730},
year = {2020}
}

@article{LOESS1979,
author = {William S. Cleveland},
title = {Robust Locally Weighted Regression and Smoothing Scatterplots},
journal = {Journal of the American Statistical Association},
volume = {74},
number = {368},
pages = {829--836},
year = {1979},
publisher = {Taylor \& Francis},
doi = {10.1080/01621459.1979.10481038}
}

@INPROCEEDINGS{Colalillo2022,
       author = {{Colalillo}, R. and {Pierre Auger Collaboration}},
        title = "{Downward Terrestrial Gamma-ray Flashes at the Pierre Auger Observatory?}",
    booktitle = {37th International Cosmic Ray Conference 2021},
         year = 2022,
        month = mar,
          eid = {395},
        pages = {395},
          doi = {10.22323/1.395.0395},
       adsurl = {https://ui.adsabs.harvard.edu/abs/2022icrc.confE.395C}
}

@ARTICLE{2020BelzUtahTAB,
       author = {{Belz}, J.~W. and {Krehbiel}, P.~R. and {Remington}, J. and {Stanley}, M.~A. and {Abbasi}, R.~U. and {LeVon}, R. and {Rison}, W. and {Rodeheffer}, D. and {Abu-Zayyad}, T. and {Allen}, M. and {Barcikowski}, E. and {Bergman}, D.~R. and {Blake}, S.~A. and {Byrne}, M. and {Cady}, R. and {Cheon}, B.~G. and {Chikawa}, M. and {di Matteo}, A. and {Fujii}, T. and {Fujita}, K. and {Fujiwara}, R. and {Fukushima}, M. and {Furlich}, G. and {Hanlon}, W. and {Hayashi}, M. and {Hayashi}, Y. and {Hayashida}, N. and {Hibino}, K. and {Honda}, K. and {Ikeda}, D. and {Inadomi}, T. and {Inoue}, N. and {Ishii}, T. and {Ito}, H. and {Ivanov}, D. and {Iwakura}, H. and {Jeong}, H.~M. and {Jeong}, S. and {Jui}, C.~C.~H. and {Kadota}, K. and {Kakimoto}, F. and {Kalashev}, O. and {Kasahara}, K. and {Kasami}, S. and {Kawai}, H. and {Kawakami}, S. and {Kawata}, K. and {Kido}, E. and {Kim}, H.~B. and {Kim}, J.~H. and {Kim}, J.~H. and {Kuzmin'}, V. and {Kuznetsov}, M. and {Kwon}, Y.~J. and {Lee}, K.~H. and {Lubsandorzhiev}, B. and {Lundquist}, J.~P. and {Machida}, K. and {Matsumiya}, H. and {Matthews}, J.~N. and {Matuyama}, T. and {Mayta}, R. and {Minamino}, M. and {Mukai}, K. and {Myers}, I. and {Nagataki}, S. and {Nakai}, K. and {Nakamura}, R. and {Nakamura}, T. and {Nakamura}, Y. and {Nonaka}, T. and {Oda}, H. and {Ogio}, S. and {Ohnishi}, M. and {Ohoka}, H. and {Oku}, Y. and {Okuda}, T. and {Omura}, Y. and {Ono}, M. and {Oshima}, A. and {Ozawa}, S. and {Park}, I.~H. and {Potts}, M. and {Pshirkov}, M.~S. and {Rodriguez}, D.~C. and {Rubtsov}, G. and {Ryu}, D. and {Sagawa}, H. and {Sahara}, R. and {Saito}, K. and {Saito}, Y. and {Sakaki}, N. and {Sako}, T. and {Sakurai}, N. and {Sano}, K. and {Seki}, T. and {Sekino}, K. and {Shibata}, F. and {Shibata}, T. and {Shimodaira}, H. and {Shin}, B.~K. and {Shin}, H.~S. and {Smith}, J.~D. and {Sokolsky}, P. and {Sone}, N. and {Stokes}, B.~T. and {Stroman}, T.~A. and {Takagi}, Y. and {Takahashi}, Y. and {Takeda}, M. and {Takeishi}, R. and {Taketa}, A. and {Takita}, M. and {Tameda}, Y. and {Tanaka}, K. and {Tanaka}, M. and {Tanoue}, Y. and {Thomas}, S.~B. and {Thomson}, G.~B. and {Tinyakov}, P. and {Tkachev}, I. and {Tokuno}, H. and {Tomida}, T. and {Troitsky}, S. and {Tsunesada}, Y. and {Uchihori}, Y. and {Udo}, S. and {Uehama}, T. and {Urban}, F. and {Wallace}, M. and {Wong}, T. and {Yamamoto}, M. and {Yamaoka}, H. and {Yamazaki}, K. and {Yashiro}, K. and {Yosei}, M. and {Yoshii}, H. and {Zhezher}, Y. and {Zundel}, Z.},
        title = "{Observations of the Origin of Downward Terrestrial Gamma-Ray Flashes}",
      journal = {Journal of Geophysical Research (Atmospheres)},
         year = 2020,
        month = dec,
       volume = {125},
       number = {23},
          eid = {e31940},
        pages = {e31940},
          doi = {10.1029/2019JD031940},
archivePrefix = {arXiv},
       eprint = {2009.14327},
 primaryClass = {physics.ao-ph},
       adsurl = {https://ui.adsabs.harvard.edu/abs/2020JGRD..12531940B}
}

@article{Ortberg23,
author = {Ortberg, John and Smith, David M. and Kamogawa, Masashi and Dwyer, Joseph and Bowers, Gregory and Chaffin, Jeffrey and Lapierre, Jeff and Wang, Daohong and Wu, Ting and Suzuki, Tomoyuki},
title = {Two Laterally Distant TGFs From Negative Cloud-To-Ground Strokes in Uchinada, Japan},
journal = {Journal of Geophysical Research: Atmospheres},
volume = {129},
number = {2},
pages = {e2023JD039020},
doi = {https://doi.org/10.1029/2023JD039020},
url = {https://agupubs.onlinelibrary.wiley.com/doi/abs/10.1029/2023JD039020},
eprint = {https://agupubs.onlinelibrary.wiley.com/doi/pdf/10.1029/2023JD039020},
note = {e2023JD039020 2023JD039020},
year = {2024}
}

@article{dwyer04rocket,
author = {Dwyer, J. R. and Rassoul, H. K. and Al-Dayeh, M. and Caraway, L. and Wright, B. and Chrest, A. and Uman, M. A. and Rakov, V. A. and Rambo, K. J. and Jordan, D. M. and Jerauld, J. and Smyth, C.},
title = {A ground level gamma-ray burst observed in association with rocket-triggered lightning},
journal = {Geophysical Research Letters},
volume = {31},
number = {5},
pages = {},
doi = {https://doi.org/10.1029/2003GL018771},
url = {https://agupubs.onlinelibrary.wiley.com/doi/abs/10.1029/2003GL018771},
eprint = {https://agupubs.onlinelibrary.wiley.com/doi/pdf/10.1029/2003GL018771},
year = {2004}
}

@ARTICLE{geantCite2,
  author={Allison, J. and Amako, K. and Apostolakis, J. and Araujo, H. and Arce Dubois, P. and Asai, M. and Barrand, G. and Capra, R. and Chauvie, S. and Chytracek, R. and Cirrone, G.A.P. and Cooperman, G. and Cosmo, G. and Cuttone, G. and Daquino, G.G. and Donszelmann, M. and Dressel, M. and Folger, G. and Foppiano, F. and Generowicz, J. and Grichine, V. and Guatelli, S. and Gumplinger, P. and Heikkinen, A. and Hrivnacova, I. and Howard, A. and Incerti, S. and Ivanchenko, V. and Johnson, T. and Jones, F. and Koi, T. and Kokoulin, R. and Kossov, M. and Kurashige, H. and Lara, V. and Larsson, S. and Lei, F. and Link, O. and Longo, F. and Maire, M. and Mantero, A. and Mascialino, B. and McLaren, I. and Mendez Lorenzo, P. and Minamimoto, K. and Murakami, K. and Nieminen, P. and Pandola, L. and Parlati, S. and Peralta, L. and Perl, J. and Pfeiffer, A. and Pia, M.G. and Ribon, A. and Rodrigues, P. and Russo, G. and Sadilov, S. and Santin, G. and Sasaki, T. and Smith, D. and Starkov, N. and Tanaka, S. and Tcherniaev, E. and Tome, B. and Trindade, A. and Truscott, P. and Urban, L. and Verderi, M. and Walkden, A. and Wellisch, J.P. and Williams, D.C. and Wright, D. and Yoshida, H.},
  journal={IEEE Transactions on Nuclear Science}, 
  title={Geant4 developments and applications}, 
  year={2006},
  volume={53},
  number={1},
  pages={270-278},
  doi={10.1109/TNS.2006.869826}}

@article{dwyer03limit,
	author={J. R. Dwyer},
	title={A fundamental limit on electric fields in air},
	journal = grl,
	volume={30},
        number={20},
	pages ={2055},
	year={2003}
}

@article{geant2016,
title = "Recent developments in Geant4",
journal = "Nuclear Instruments and Methods in Physics Research Section A: Accelerators, Spectrometers, Detectors and Associated Equipment",
volume = "835",
pages = "186 - 225",
year = "2016",
issn = "0168-9002",
doi = "https://doi.org/10.1016/j.nima.2016.06.125",
url = "http://www.sciencedirect.com/science/article/pii/S0168900216306957",
author = {J. Allison and K. Amako and J. Apostolakis and P. Arce and M. Asai and T. Aso and E. Bagli and A. Bagulya and S. Banerjee and G. Barrand and B.R. Beck and A.G. Bogdanov and D. Brandt and J.M.C. Brown and H. Burkhardt and Ph. Canal and D. Cano-Ott and S. Chauvie and K. Cho and G.A.P. Cirrone and G. Cooperman and M.A. Cortés-Giraldo and G. Cosmo and G. Cuttone and G. Depaola and L. Desorgher and X. Dong and A. Dotti and V.D. Elvira and G. Folger and Z. Francis and A. Galoyan and L. Garnier and M. Gayer and K.L. Genser and V.M. Grichine and S. Guatelli and P. Guèye and P. Gumplinger and A.S. Howard and I. Hřivnáčová and S. Hwang and S. Incerti and A. Ivanchenko and V.N. Ivanchenko and F.W. Jones and S.Y. Jun and P. Kaitaniemi and N. Karakatsanis and M. Karamitros and M. Kelsey and A. Kimura and T. Koi and H. Kurashige and A. Lechner and S.B. Lee and F. Longo and M. Maire and D. Mancusi and A. Mantero and E. Mendoza and B. Morgan and K. Murakami and T. Nikitina and L. Pandola and P. Paprocki and J. Perl and I. Petrović and M.G. Pia and W. Pokorski and J.M. Quesada and M. Raine and M.A. Reis and A. Ribon and A. Ristić Fira and F. Romano and G. Russo and G. Santin and T. Sasaki and D. Sawkey and J.I. Shin and I.I. Strakovsky and A. Taborda and S. Tanaka and B. Tomé and T. Toshito and H.N. Tran and P.R. Truscott and L. Urban and V. Uzhinsky and J.M. Verbeke and M. Verderi and B.L. Wendt and H. Wenzel and D.H. Wright and D.M. Wright and T. Yamashita and J. Yarba and H. Yoshida}
}

@TECHREPORT{UsStdAtm76,
        title = "{U.S. standard atmosphere}",
    booktitle = {Unknown},
         year = 1976,
        month = Oct,
       adsurl = {https://ui.adsabs.harvard.edu/#abs/1976ussa.rept......},
      institution = {Unknown},
      address = {Unknown}
}

@ARTICLE{AgostinelliGeant4,
       author = {{Agostinelli}, S. and {Allison}, J. and {Amako}, K. and {Apostolakis},
        J. and {Araujo}, H. and {Arce}, P. and {Asai}, M. and {Axen}, D.
        and {Banerjee}, S. and {Barrand}, G. and {Behner}, F. and
        {Bellagamba}, L. and {Boudreau}, J. and {Broglia}, L. and
        {Brunengo}, A. and {Burkhardt}, H. and {Chauvie}, S. and
        {Chuma}, J. and {Chytracek}, R. and {Cooperman}, G. and {Cosmo},
        G. and {Degtyarenko}, P. and {Dell'Acqua}, A. and {Depaola}, G.
        and {Dietrich}, D. and {Enami}, R. and {Feliciello}, A. and
        {Ferguson}, C. and {Fesefeldt}, H. and {Folger}, G. and
        {Foppiano}, F. and {Forti}, A. and {Garelli}, S. and {Giani}, S.
        and {Giannitrapani}, R. and {Gibin}, D. and {G{\'o}mez Cadenas},
        J.~J. and {Gonz{\'a}lez}, I. and {Gracia Abril}, G. and
        {Greeniaus}, G. and {Greiner}, W. and {Grichine}, V. and
        {Grossheim}, A. and {Guatelli}, S. and {Gumplinger}, P. and
        {Hamatsu}, R. and {Hashimoto}, K. and {Hasui}, H. and
        {Heikkinen}, A. and {Howard}, A. and {Ivanchenko}, V. and
        {Johnson}, A. and {Jones}, F.~W. and {Kallenbach}, J. and
        {Kanaya}, N. and {Kawabata}, M. and {Kawabata}, Y. and
        {Kawaguti}, M. and {Kelner}, S. and {Kent}, P. and {Kimura}, A.
        and {Kodama}, T. and {Kokoulin}, R. and {Kossov}, M. and
        {Kurashige}, H. and {Lamanna}, E. and {Lamp{\'e}n}, T. and
        {Lara}, V. and {Lefebure}, V. and {Lei}, F. and {Liendl}, M. and
        {Lockman}, W. and {Longo}, F. and {Magni}, S. and {Maire}, M.
        and {Medernach}, E. and {Minamimoto}, K. and {Mora de Freitas},
        P. and {Morita}, Y. and {Murakami}, K. and {Nagamatu}, M. and
        {Nartallo}, R. and {Nieminen}, P. and {Nishimura}, T. and
        {Ohtsubo}, K. and {Okamura}, M. and {O'Neale}, S. and {Oohata},
        Y. and {Paech}, K. and {Perl}, J. and {Pfeiffer}, A. and {Pia},
        M.~G. and {Ranjard}, F. and {Rybin}, A. and {Sadilov}, S. and
        {Di Salvo}, E. and {Santin}, G. and {Sasaki}, T. and {Savvas},
        N. and {Sawada}, Y. and {Scherer}, S. and {Sei}, S. and
        {Sirotenko}, V. and {Smith}, D. and {Starkov}, N. and
        {Stoecker}, H. and {Sulkimo}, J. and {Takahata}, M. and
        {Tanaka}, S. and {Tcherniaev}, E. and {Safai Tehrani}, E. and
        {Tropeano}, M. and {Truscott}, P. and {Uno}, H. and {Urban}, L.
        and {Urban}, P. and {Verderi}, M. and {Walkden}, A. and
        {Wander}, W. and {Weber}, H. and {Wellisch}, J.~P. and {Wenaus},
        T. and {Williams}, D.~C. and {Wright}, D. and {Yamada}, T. and
        {Yoshida}, H. and {Zschiesche}, D. and {G EANT4 Collaboration}},
        title = "{G EANT4{\textemdash}a simulation toolkit}",
      journal = {Nuclear Instruments and Methods in Physics Research A},
         year = 2003,
        month = Jul,
       volume = {506},
        pages = {250-303},
          doi = {10.1016/S0168-9002(03)01368-8},
       adsurl = {https://ui.adsabs.harvard.edu/#abs/2003NIMPA.506..250A}
}

@article{ABRAHAM201029,
title = {Trigger and aperture of the surface detector array of the Pierre Auger Observatory},
journal = {Nucl. Instrum. Methods A},
volume = {613},
number = {1},
pages = {29--39},
year = {2010},
issn = {0168-9002},
doi = {https://doi.org/10.1016/j.nima.2009.11.018},
url = {https://www.sciencedirect.com/science/article/pii/S0168900209021688},
;author = {J. Abraham and others},
author = {{Pierre Auger Collaboration}},
collaboration = {Pierre Auger},
}

@article{auger_dipole_paper,
title = {Observation of a Large-scale Anisotropy in the Arrival Directions of Cosmic Rays above $8\times10^18$\,eV},
journal = {Science},
volume = {357},
pages = {1266 -- 1270},
year = {2017},
doi = {https://doi.org/10.1126/science.aan4338},
url = {https://www.science.org/doi/10.1126/science.aan4338},
author = {{Pierre Auger Collaboration}},
collaboration = {Pierre Auger},
}
\clearpage

\small\selectfont\setlength{\baselineskip}{0.8\baselineskip}\justifying
\begin{wrapfigure}[9]{l}{0.12\linewidth}
\includegraphics[width=0.98\linewidth]{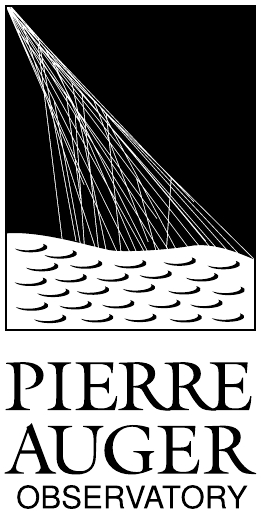}
\end{wrapfigure}
\begin{sloppypar}\noindent
A.~Abdul Halim$^{13}$,
P.~Abreu$^{67}$,
M.~Aglietta$^{50,49}$,
M.~Ahmed$^{31}$,
I.~Allekotte$^{1}$,
K.~Almeida Cheminant$^{75,74}$,
R.~Aloisio$^{42,43}$,
J.~Alvarez-Mu\~niz$^{73}$,
A.~Ambrosone$^{42,43}$,
J.~Ammerman Yebra$^{73}$
% ...

\end{sloppypar}
\begin{center}
\par\noindent
\textbf{The Pierre Auger Collaboration}
\end{center}

\vspace{1ex}
\begin{center}
\rule{0.1\columnwidth}{0.5pt}
\raisebox{-0.4ex}{\scriptsize$\bullet$}
\rule{0.1\columnwidth}{0.5pt}
\end{center}

\vspace{1ex}
\begin{description}[labelsep=0.2em,align=right,labelwidth=0.7em,labelindent=0em,leftmargin=2em,noitemsep,before={\renewcommand\makelabel[1]{##1 }}]
\item[$^{1}$] Centro At\'omico Bariloche and Instituto Balseiro (CNEA-UNCuyo-CONICET), San Carlos de Bariloche, Argentina
\item[$^{2}$] Departamento de F\'\i{}sica and Departamento de Ciencias de la Atm\'osfera y los Oc\'eanos, FCEyN, Universidad de Buenos Aires and CONICET, Buenos Aires, Argentina
\item[$^{3}$] IFLP, Universidad Nacional de La Plata and CONICET, La Plata, Argentina
\item[$^{4}$] Instituto de Astronom\'\i{}a y F\'\i{}sica del Espacio (IAFE, CONICET-UBA), Buenos Aires, Argentina
\item[$^{5}$] Instituto de F\'\i{}sica de Rosario (IFIR) -- CONICET/U.N.R.\ and Facultad de Ciencias Bioqu\'\i{}micas y Farmac\'euticas U.N.R., Rosario, Argentina
\item[$^{6}$] Instituto de Tecnolog\'\i{}as en Detecci\'on y Astropart\'\i{}culas (CNEA, CONICET, UNSAM), and Universidad Tecnol\'ogica Nacional -- Facultad Regional Mendoza (CONICET/CNEA), Mendoza, Argentina
\item[$^{7}$] Instituto de Tecnolog\'\i{}as en Detecci\'on y Astropart\'\i{}culas (CNEA, CONICET, UNSAM), Buenos Aires, Argentina
\item[$^{8}$] International Center of Advanced Studies and Instituto de Ciencias F\'\i{}sicas, ECyT-UNSAM and CONICET, Campus Miguelete -- San Mart\'\i{}n, Buenos Aires, Argentina
\item[$^{9}$] Laboratorio Atm\'osfera -- Departamento de Investigaciones en L\'aseres y sus Aplicaciones -- UNIDEF (CITEDEF-CONICET), Argentina
\item[$^{10}$] Observatorio Pierre Auger, Malarg\"ue, Argentina
\item[$^{11}$] Observatorio Pierre Auger and Comisi\'on Nacional de Energ\'\i{}a At\'omica, Malarg\"ue, Argentina
\item[$^{12}$] Universidad Tecnol\'ogica Nacional -- Facultad Regional Buenos Aires, Buenos Aires, Argentina
\item[$^{13}$] Adelaide University, Adelaide, S.A., Australia
\item[$^{14}$] Universit\'e Libre de Bruxelles (ULB), Brussels, Belgium
\item[$^{15}$] Vrije Universiteit Brussels, Brussels, Belgium
\item[$^{16}$] Centro Brasileiro de Pesquisas Fisicas, Rio de Janeiro, RJ, Brazil
\item[$^{17}$] Centro Federal de Educa\c{c}\~ao Tecnol\'ogica Celso Suckow da Fonseca, Petropolis, Brazil
\item[$^{18}$] Universidade de S\~ao Paulo, Escola de Engenharia de Lorena, Lorena, SP, Brazil
\item[$^{19}$] Universidade de S\~ao Paulo, Instituto de F\'\i{}sica de S\~ao Carlos, S\~ao Carlos, SP, Brazil
\item[$^{20}$] Universidade de S\~ao Paulo, Instituto de F\'\i{}sica, S\~ao Paulo, SP, Brazil
\item[$^{21}$] Universidade Estadual de Campinas (UNICAMP), IFGW, Campinas, SP, Brazil
\item[$^{22}$] Universidade Estadual de Feira de Santana, Feira de Santana, Brazil
\item[$^{23}$] Universidade Federal do ABC, Santo Andr\'e, SP, Brazil
\item[$^{24}$] Universidade Federal do Paran\'a, Setor Palotina, Palotina, Brazil
\item[$^{25}$] Universidade Federal do Rio de Janeiro, Instituto de F\'\i{}sica, Rio de Janeiro, RJ, Brazil
\item[$^{26}$] Universidad de Medell\'\i{}n, Medell\'\i{}n, Colombia
\item[$^{27}$] Universidad Industrial de Santander, Bucaramanga, Colombia
\item[$^{28}$] Charles University, Faculty of Mathematics and Physics, Institute of Particle and Nuclear Physics, Prague, Czech Republic
\item[$^{29}$] Institute of Physics of the Czech Academy of Sciences, Prague, Czech Republic
\item[$^{30}$] Palacky University, Olomouc, Czech Republic
\item[$^{31}$] CNRS/IN2P3, IJCLab, Universit\'e Paris-Saclay, Orsay, France
\item[$^{32}$] Laboratoire de Physique Nucl\'eaire et de Hautes Energies (LPNHE), Sorbonne Universit\'e, Universit\'e de Paris, CNRS-IN2P3, Paris, France
\item[$^{33}$] Universit\'e Paris-Saclay, CNRS/IN2P3, IJCLab, Orsay, France
\item[$^{34}$] Bergische Universit\"at Wuppertal, Department of Physics, Wuppertal, Germany
\item[$^{35}$] Karlsruhe Institute of Technology (KIT), Institute for Experimental Particle Physics, Karlsruhe, Germany
\item[$^{36}$] Karlsruhe Institute of Technology (KIT), Institut f\"ur Prozessdatenverarbeitung und Elektronik, Karlsruhe, Germany
\item[$^{37}$] Karlsruhe Institute of Technology (KIT), Institute for Astroparticle Physics, Karlsruhe, Germany
\item[$^{38}$] RWTH Aachen University, III.\ Physikalisches Institut A, Aachen, Germany
\item[$^{39}$] TU Dortmund University, Department of Physics, Dortmund, Germany
\item[$^{40}$] Universit\"at Hamburg, II.\ Institut f\"ur Theoretische Physik, Hamburg, Germany
\item[$^{41}$] Universit\"at Siegen, Department Physik -- Experimentelle Teilchenphysik, Siegen, Germany
\item[$^{42}$] Gran Sasso Science Institute, L'Aquila, Italy
\item[$^{43}$] INFN Laboratori Nazionali del Gran Sasso, Assergi (L'Aquila), Italy
\item[$^{44}$] INFN, Sezione di Catania, Catania, Italy
\item[$^{45}$] INFN, Sezione di Lecce, Lecce, Italy
\item[$^{46}$] INFN, Sezione di Milano, Milano, Italy
\item[$^{47}$] INFN, Sezione di Napoli, Napoli, Italy
\item[$^{48}$] INFN, Sezione di Roma ``Tor Vergata'', Roma, Italy
\item[$^{49}$] INFN, Sezione di Torino, Torino, Italy
\item[$^{50}$] Osservatorio Astrofisico di Torino (INAF), Torino, Italy
\item[$^{51}$] Politecnico di Milano, Dipartimento di Scienze e Tecnologie Aerospaziali , Milano, Italy
\item[$^{52}$] Universit\`a del Salento, Dipartimento di Matematica e Fisica ``E.\ De Giorgi'', Lecce, Italy
\item[$^{53}$] Universit\`a dell'Aquila, Dipartimento di Scienze Fisiche e Chimiche, L'Aquila, Italy
\item[$^{54}$] Universit\`a di Catania, Dipartimento di Fisica e Astronomia ``Ettore Majorana``, Catania, Italy
\item[$^{55}$] Universit\`a di Milano, Dipartimento di Fisica, Milano, Italy
\item[$^{56}$] Universit\`a di Napoli ``Federico II'', Dipartimento di Fisica ``Ettore Pancini'', Napoli, Italy
\item[$^{57}$] Universit\`a di Palermo, Dipartimento di Fisica e Chimica ''E.\ Segr\`e'', Palermo, Italy
\item[$^{58}$] Universit\`a di Roma ``Tor Vergata'', Dipartimento di Fisica, Roma, Italy
\item[$^{59}$] Universit\`a Torino, Dipartimento di Fisica, Torino, Italy
\item[$^{60}$] Benem\'erita Universidad Aut\'onoma de Puebla, Puebla, M\'exico
\item[$^{61}$] Unidad Profesional Interdisciplinaria en Ingenier\'\i{}a y Tecnolog\'\i{}as Avanzadas del Instituto Polit\'ecnico Nacional (UPIITA-IPN), M\'exico, D.F., M\'exico
\item[$^{62}$] Universidad Aut\'onoma de Chiapas, Tuxtla Guti\'errez, Chiapas, M\'exico
\item[$^{63}$] Universidad Michoacana de San Nicol\'as de Hidalgo, Morelia, Michoac\'an, M\'exico
\item[$^{64}$] Universidad Nacional Aut\'onoma de M\'exico, M\'exico, D.F., M\'exico
\item[$^{65}$] Institute of Nuclear Physics PAN, Krakow, Poland
\item[$^{66}$] University of \L{}\'od\'z, Faculty of High-Energy Astrophysics,\L{}\'od\'z, Poland
\item[$^{67}$] Laborat\'orio de Instrumenta\c{c}\~ao e F\'\i{}sica Experimental de Part\'\i{}culas -- LIP and Instituto Superior T\'ecnico -- IST, Universidade de Lisboa -- UL, Lisboa, Portugal
\item[$^{68}$] ``Horia Hulubei'' National Institute for Physics and Nuclear Engineering, Bucharest-Magurele, Romania
\item[$^{69}$] Institute of Space Science, Bucharest-Magurele, Romania
\item[$^{70}$] Center for Astrophysics and Cosmology (CAC), University of Nova Gorica, Nova Gorica, Slovenia
\item[$^{71}$] Experimental Particle Physics Department, J.\ Stefan Institute, Ljubljana, Slovenia
\item[$^{72}$] Universidad de Granada and C.A.F.P.E., Granada, Spain
\item[$^{73}$] Instituto Galego de F\'\i{}sica de Altas Enerx\'\i{}as (IGFAE), Universidade de Santiago de Compostela, Santiago de Compostela, Spain
\item[$^{74}$] IMAPP, Radboud University Nijmegen, Nijmegen, The Netherlands
\item[$^{75}$] Nationaal Instituut voor Kernfysica en Hoge Energie Fysica (NIKHEF), Science Park, Amsterdam, The Netherlands
\item[$^{76}$] Stichting Astronomisch Onderzoek in Nederland (ASTRON), Dwingeloo, The Netherlands
\item[$^{77}$] Case Western Reserve University, Cleveland, OH, USA
\item[$^{78}$] Colorado School of Mines, Golden, CO, USA
\item[$^{79}$] Department of Physics and Astronomy, Lehman College, City University of New York, Bronx, NY, USA
\item[$^{80}$] Michigan Technological University, Houghton, MI, USA
\item[$^{81}$] New York University, New York, NY, USA
\item[$^{82}$] University of Chicago, Enrico Fermi Institute, Chicago, IL, USA
\item[$^{83}$] University of Delaware, Department of Physics and Astronomy, Bartol Research Institute, Newark, DE, USA
\item[] -----
\item[$^{a}$] Max-Planck-Institut f\"ur Radioastronomie, Bonn, Germany
\item[$^{b}$] also at Kapteyn Institute, University of Groningen, Groningen, The Netherlands
\item[$^{c}$] School of Physics and Astronomy, University of Leeds, Leeds, United Kingdom
\item[$^{d}$] Fermi National Accelerator Laboratory, Fermilab, Batavia, IL, USA (Affiliation for identification purposes only)
\item[$^{e}$] Pennsylvania State University, University Park, PA, USA
\item[$^{f}$] Colorado State University, Fort Collins, CO, USA
\item[$^{g}$] Louisiana State University, Baton Rouge, LA, USA
\item[$^{h}$] Universit\'e Grenoble Alpes, CNRS, Grenoble Institute of Engineering, LPSC-IN2P3, Grenoble, France
\item[$^{i}$] now at Graduate School of Science, Osaka Metropolitan University, Osaka, Japan
\item[$^{j}$] Institut universitaire de France (IUF), Paris, France
\item[$^{k}$] now at Fraunhofer Institute for Integrated Circuits (IIS), Erlangen, Germany
\end{description}

\end{document}